\documentclass[lettersize,journal]{IEEEtran}
\usepackage{amsmath,amsfonts}
\usepackage{algorithmic}
\usepackage{algorithm}
\usepackage{array}
\usepackage{textcomp}
\usepackage{stfloats}
\usepackage{url}
\usepackage{verbatim}
\usepackage{graphicx}
\usepackage{caption}
\usepackage{subfigure}
\usepackage{cite}
\usepackage{amssymb}
\usepackage{amsthm} % proof 所需包
\usepackage{color}
\usepackage{hyperref}
\begin{document}

\title{Pattern Zooming: Near-Field Wideband Beam Training with Wavenumber-Domain Codebook}

\author{Caihao Weng, Ying Wang,~\IEEEmembership{Member,~IEEE}, Xufeng Guo, Yuqing Guo, and Ce Guo
        % <-this % stops a space
\thanks{Caihao Weng, Ying Wang, Xufeng Guo, Yuqing Guo, and Ce Guo are with the State Key Laboratory of Networking and Switching Technology, Beijing University of
Posts and Telecommunications, Beijing, China 100876 (e-mail: wengcaihao@bupt.edu.cn; wangying@bupt.edu.cn; brook1711@bupt.edu.cn; guoyuqing2020@bupt.edu.cn; ceguo@bupt.edu.cn).}% <-this % stops a space
}

\markboth{Journal of \LaTeX\ Class Files,~Vol.~14, No.~8, August~2021}%
{Shell \MakeLowercase{\textit{et al.}}: A Sample Article Using IEEEtran.cls for IEEE Journals}


\maketitle

\begin{abstract}
Near-field beam training is essential for harvesting the high-gain potential of extremely large-scale multiple input multiple output systems. To reduce training overhead, existing works have mostly leveraged the beam squint effect by utilizing time-delay (TD) beamforming with polar-domain codebook, which enables the simultaneous sweeping of multiple angles at a specific distance. However, such methods still suffer from high overhead due to the exhaustive distance searching. To address this challenge, we propose a pattern zooming based near-field wideband beam training with wavenumber-domain codebook. Specifically, we first establish a refined wideband Fourier plane-wave channel representation, based on which we reveal a pattern zooming effect, where the wavenumber-domain patterns across different subcarriers exhibit frequency-dependent scaling relative to the center frequency. Exploiting this property, we develop a TD-assisted beam sweeping strategy that simultaneously probes multiple wavenumber directions to rapidly acquire the complete wavenumber-domain pattern. Based on the acquired pattern, we further derive exact, approximation-free closed-form expressions that establish a rigorous mapping between the receiver coordinates and the wavenumber-domain pattern, enabling accurate user localization with only a few pilots. Finally, numerical results validate the superiority of our proposed scheme in terms of both beamforming gain and training overhead.

\end{abstract} 


\section{Introduction}
Future sixth-generation (6G) wireless communications are expected to undergo a paradigm shift toward ubiquitous connectivity, targeting peak data rates of 1 terabit per second (Tb/s) and sub-millisecond over-the-air latency~\cite{6G-1, NF-tutorial-1}. This pursuit of extreme throughput drives the migration to higher frequency bands, such as millimeter-wave (mmWave) and terahertz (THz), to exploit their abundant spectrum resources\cite{mmWave-1}. Benefiting from this migration, extremely large antenna arrays (ELAAs) have been proposed as a pivotal technology for 6G\cite{NF-tutorial-2,NF-tutorial-3,NF-tutorial-4}. On one hand, the shortened wavelengths at these higher frequencies facilitate the practical deployment of ELAAs with thousands of elements, while the abundant spectrum provides the massive bandwidth necessary for extremely large-scale multiple-input-multiple-output (XL-MIMO) systems\cite{Tcom-DLL}. On the other hand, XL-MIMO provides the essential high-gain beamforming to compensate for the severe path loss inherent in high-frequency propagation\cite{rainbow}. Driven by this synergy, wideband XL-MIMO has emerged as a cornerstone candidate for future wireless communications \cite{rainbow,wideband_beamforming,wide_beamforming_lyw}.

Beyond the substantial increase in antenna count, the most profound impact of XL-MIMO lies in the transformation of the electromagnetic (EM) propagation environment. Specifically, the EM radiation field is typically partitioned into far-field and near-field regions, with the Rayleigh distance serving as the boundary\cite{Rayleigh, boundary_3}. In the far-field region, EM waves are conventionally modeled as planar waves, enabling the system to steer signals toward specific angular directions, i.e., beam-steering. Conversely, in the near-field region, the planar-wave model becomes inadequate. Instead, EM waves must be characterized by spherical wavefronts, which facilitate spatial focusing to concentrate energy at a specific location rather than a mere direction, i.e., beam-focusing\cite{Tcom-DLL}. Compared to traditional beam-steering, this capability provides enhanced array gain and additional degrees of freedom. As the physical aperture of ELAAs expands, the Rayleigh distance extends significantly to encompass typical communication ranges, thereby necessitating that future XL-MIMO systems operate in near-field or hybrid far and near-field environments\cite{wch-twc}. To harvest the high-gain potential inherent in near-field focusing, efficient beam training is indispensable for establishing precise spatial links\cite{wideband_bt}. During this process, the transmitter (Tx) evaluates codewords from a predefined codebook to identify the optimal entry that maximizes the beamforming gain\cite{wch-twc}. Consequently, minimizing the training overhead while maintaining high beamforming gain remains a fundamental and persistent challenge in beam training design, which has attracted extensive research interest.

\subsection{Prior Works}
To address the training overhead challenge in XL-MIMO, existing research has evolved along two distinct paths, tailored to the specific characteristics of far-field and near-field regions, respectively. For far-field communication, discrete Fourier transform (DFT) codebooks are widely utilized to identify the optimal beam direction. While exhaustive sweeping through such codebooks is viable for small-scale arrays, it becomes prohibitively expensive for XL-MIMO systems, as the overhead scales linearly with the number of antenna elements~\cite{NF-tutorial-1}. To mitigate this, hierarchical beam training frameworks have been developed, employing a coarse-to-fine search mechanism across codebooks of varying resolutions to achieve a logarithmic reduction in overhead~\cite{far-hierarchical}. Nevertheless, the transition from planar to spherical wavefronts introduces an additional distance dimension to the search space. Consequently, earlier works generally assumed that DFT codebooks would suffer from severe performance degradation in near-field scenarios due to their lack of distance-domain sweeping capability.

Based on this belief, dedicated near-field beam training schemes have been proposed by incorporating the distance dimension into the codebook design. Specifically, polar-domain codebooks were introduced in~\cite{narrow_near} to sample the spatial domain jointly over angle and distance, thereby accommodating the spherical-wave characteristics of near-field propagation. However, the two-dimensional (2D) search space leads to substantially increased training overhead. To address this issue, various low-overhead near-field beam training schemes have been developed. Representative approaches include two-stage search strategies that first identify a coarse angular sector and then refine the precise angle and distance information~\cite{near-hierarchical-1,wch-TVT}, as well as hierarchical beam training schemes based on multi-resolution polar-domain codebooks~\cite{near-hierarchical-2}.

In summary, the above works predominantly follow a divide-and-conquer paradigm, where beam training methods are specifically tailored for either far-field or near-field scenarios. However, such a segregated approach necessitates prior knowledge of the communication region, which imposes additional system overhead. Furthermore, the inclusion of the distance dimension inevitably elevates either the training overhead or the computational complexity of these near-field specific methods. These limitations motivate a critical question: \emph{Can traditional, low-overhead far-field DFT codebooks be repurposed for near-field beam training to achieve a more unified and efficient architecture across all field regions?}

Recent studies have begun to explore this possibility, revealing that far-field DFT beam patterns implicitly encapsulate information regarding the receiver (Rx) location. As an initial attempt, learning-based methods were developed to extract these complex relationships by employing neural networks as black-box estimators, without identifying specific physical features\cite{learning-1,wch-TVT}. To bridge the gap between pure black-box learning and physical interpretability, a hybrid approach has been introduced, integrating model-based and data-driven techniques for robust beam alignment~\cite{dft_bt_4}. Beyond these approaches that rely on neural networks, recent studies have extracted user location features from the width and center of the DFT beam pattern, achieving distance estimation either through analytical closed-form expressions \cite{dft_bt_2, dft_bt_3} or via one-dimensional (1D) search over the range domain \cite{dft_bt_1}. However, these pioneering attempts primarily rely on heuristic approximations or simplified models. Consequently, the exact mathematical mapping between these specific beam characteristics and receiver locations remains an open challenge.

More importantly, while the above studies provide valuable insights into unified beam training, they are exclusively tailored for narrowband systems and fail to account for the frequency-dependent nature of wideband channels, which constitute the foundation of most practical communication systems. In practice, the use of frequency-independent phase shifters (PSs) for analog beamforming creates a fundamental mismatch with the frequency-dependent characteristics of wideband systems \cite{wideband_PS}. This inconsistency triggers beam squint effect, whereby beams at different subcarriers diverge from the intended spatial location~\cite{split_1,split_2}. Specifically, in conventional far-field wideband systems employing plane-wave transmission, such as in traditional far-field beam training~\cite{far-hierarchical}, this effect manifests as the beam split effect, leading to frequency-dependent angular deviations. Conversely, in near-field wideband systems employing spherical-wave transmission, typically observed during conventional near-field beam training~\cite{narrow_near, near-hierarchical-1}, it evolves into the rainbow effect, where diverse frequencies focus at distinct angles and distances~\cite{rainbow}. Consequently, signal energy across the entire bandwidth fails to focus coherently at the target location, leading to significant beamforming gain degradation.

To address the challenges posed by beam squint, the time-delay (TD) beamforming architecture has been widely adopted in existing wideband beam training schemes for two primary reasons. First, by leveraging the frequency-dependent phase response inherent in TD beamforming, beams across the entire bandwidth can be precisely aligned, thereby eliminating the beam squint effect during transmission. Second, TD beamforming enables a controllable beam squint, which can be exploited to accelerate the training process. In contrast to the data transmission stage where beam squint is detrimental, the training stage harnesses this effect to facilitate frequency-division sweeping. This allows different subcarriers to probe diverse angular directions in far-field communications~\cite{wide_far}, or detect distinct locations in near-field scenarios~\cite{wideband_bt}. However, the beam split and rainbow effects are tied to conventional segregated beam training paradigms. As established earlier, recent trends indicate that plane-wave-based codebooks, such as DFT codebooks, can also be employed for near-field beam training to avoid the prohibitive overhead of polar-domain codebooks. Nevertheless, these unified approaches are limited to narrowband systems. It remains unexplored how the frequency-dependent characteristics of wideband systems affect plane-wave transmission when its application is extended beyond the far field into the near field. This gap motivates our investigation into the underlying wideband effect when applying plane-wave-based codebooks across all field regions.

\subsection{Contributions}
Inspired by the above, this paper generalizes the methodology of utilizing far-field codebooks for near-field beam training to accommodate practical wideband systems. To this end, we propose an efficient near-field wideband beam training scheme based on a wavenumber-domain codebook consisting of DFT vectors\footnote{While the wavenumber-domain codebook is composed of DFT vectors, it uniquely excludes evanescent wave components to reduce training overhead. Consequently, for a 2D uniform planar array (UPA), the wavenumber-domain codebook requires lower overhead than a conventional 2D DFT codebook, whereas they are equivalent for a 1D uniform linear array (ULA). Further details regarding this distinction are available in \cite{wch-twc} and \cite{wd_codebook}. Throughout this paper, the terms wavenumber-domain codebook and far-field DFT codebook are used interchangeably.}. Specifically, the main contributions of this paper are summarized as follows.

\begin{itemize}
\item[$\bullet$] First, we analyze the Fourier plane-wave (FPW) representation of the narrowband channel and draw some insights into its limitations regarding high-frequency aliasing and low-frequency redundancy in wideband systems. Accordingly, the conventional FPW model is extended to establish a refined FPW representation of the wideband channel, which provides a unified and accurate characterization of the EM propagation environment across both far-field and near-field regions.

\item[$\bullet$] Second, we unveil the pattern zooming effect that emerges when plane-wave transmission is employed across both far-field and near-field regions in wideband systems utilizing PS beamforming architectures. Specifically, we reveal that when utilizing wavenumber-domain codebooks, the acquired wavenumber-domain patterns across different subcarriers undergo a distinct frequency-dependent scaling relative to the center frequency. In particular, patterns at lower frequencies manifest as reduced versions of the central pattern, whereas those at higher frequencies appear clearly magnified.

\item[$\bullet$] Third, we demonstrate that TD beamforming can flexibly control the degree of the pattern zooming effect. By exploiting this controllability, we propose an efficient near-field wideband beam training scheme based on the wavenumber-domain codebook. Specifically, TD beamforming enables different subcarriers to simultaneously probe distinct wavenumber directions within each time slot, thereby rapidly acquiring the complete wavenumber-domain pattern. We further reveal that the receiver's spatial coordinates can be uniquely characterized by two 1D pattern features, namely the pattern width and center. Based on these features, which remain reliably extractable even under partial pattern overlap, we derive exact closed-form expressions that establish a rigorous mapping from the pattern features to the receiver's location.

\item[$\bullet$] Finally, extensive numerical results are provided to validate the effectiveness of the proposed scheme. The results demonstrate that our method achieves satisfactory rate performance while significantly reducing training overhead compared to existing benchmarks. This superiority is attributed to the enhanced localization precision enabled by the derived approximation-free closed-form expressions, which ensure robust performance across various communication scenarios.
\end{itemize}

\subsection{Organization and Notation}
\emph{Organization}: The remainder of this paper is organized as follows. Section \uppercase\expandafter{\romannumeral2} introduces the system model and reviews the boundaries of the wavenumber-domain pattern. Then, we propose the refined wideband FPW representation and introduce the pattern zooming effect in Section \uppercase\expandafter{\romannumeral3}. In Section \uppercase\expandafter{\romannumeral4}, we present the design of the proposed pattern zooming based near-field wideband beam training with the wavenumber-domain codebook. Section \uppercase\expandafter{\romannumeral5} and Section \uppercase\expandafter{\romannumeral6} provide numerical results and conclusions, respectively.

\emph{Notations}: Lowercase, bold lowercase, and bold uppercase letters denote scalars, vectors, and matrices, respectively. The symbol $\mathbb{Z}$ and $\mathbb{C}$ refer to the integer and complex numbers. The notation $\mathbf{A}_{n,:}$, $\mathbf{A}_{:,m}$, and $\mathbf{A}_{n,m}$ represent the $n$-th row of matrix $\mathbf{A}$, the $m$-th column of matrix $\mathbf{A}$, and the $\left(n,m\right)$-th entry of matrix $\mathbf{A}$. $\vert \cdot \vert$ indicates the absolute value of a scalar or the cardinality of a vector. $\mathbf{I}_N$ denotes the identity matrix of size $N \times N$.

\section{System Model}
In this section, we begin by introducing the considered wideband communication scenario. Then, the FPW representation of the narrowband channel is provided. Finally, we briefly review the boundaries of the wavenumber-domain pattern to make the paper self-contained.

\subsection{Scenario}
Consider a wideband XL-MIMO communication system where a Tx equipped with an $N$-element UPA serves a single-antenna Rx\footnote{The proposed framework can be extended to receivers equipped with a limited number of antennas. Since the one-sided near-field MIMO channel is separable~\cite{MIMO_DLL}, the transmit and receive beamformers can be designed independently. Consequently, by adopting a temporary antenna deactivation strategy that deactivates all but one receive antenna during transmit beam training~\cite{antenna_deactivation}, the proposed pattern zooming method can be directly applied.}. Without loss of generality, we assume the number of antenna elements along the $x$-axis and $y$-axis are odd, denoted by $N_x=2\tilde{N}_x + 1$ and $N_y=2\tilde{N}_y+1$, respectively, such that $N=N_x N_y$. The horizontal and vertical lengths of the UPA are $L_x=N_x\delta$ and $L_y=N_y\delta$, where $\delta=\lambda_{\text{c}}/2$ is the half-wavelength antenna spacing. To mitigate inter-symbol interference, orthogonal frequency division multiplexing (OFDM) with $M$ subcarriers is employed. Let $B$, $c$, $f_{\text{c}}$, and $\lambda_{\text{c}} = c/f_{\text{c}}$ denote the bandwidth, the speed of light, the central carrier frequency, and the central wavelength, respectively. Accordingly, the frequency of the $m$-th subcarrier is given by $f_m=f_{\text{c}}+\frac{B}{M}\left(m-1-\frac{M-1}{2}\right),\forall m=1,\ldots,M$.

\subsection{FPW Representation of the Narrowband Channel}
Given the severe path loss induced by scattering in mmWave and THz communications, this paper adopts a line-of-sight (LoS) channel model for the theoretical development of the proposed framework~\cite{dft_bt_1,rainbow}. Thus, the channel between the $n$-th antenna element of the Tx and the Rx at the $m$-th subcarrier can be expressed as
\begin{equation}
\label{Eq1_1}
\left[ \mathbf{h}_m \right]_n =\sqrt{N} \beta_{m,n} \exp \left( -jk_m r_n \right),
\end{equation}
where $n=n_yN_x + n_x$ is the antenna index, with $n_x\in \{-\tilde{N}_x,\ldots,0,\ldots,\tilde{N}_x \}$ and $n_y\in \{-\tilde{N}_y,\ldots,0,\ldots,\tilde{N}_y \}$ denoting the element indices of the Tx along the $x$- and $y$-axes, respectively. Here, $\beta_{m,n}=\lambda_m/(4\pi r_{n})$ denotes the complex gain, while $\lambda_m=c/f_m$ and $k_m=2\pi/\lambda_m$ are the wavelength and the wavenumber at subcarrier $m$. The coordinate of the Rx is $\mathbf{r}=[x_r,y_r,z_r]^T=r_u[\sin\theta_u \cos \phi_u, \sin\theta_u \sin \phi_u, \cos \theta_u]^T$, where $r_u$, $\theta_u$, and $\phi_u$ represent the distance, the elevation angle, and the azimuth angle of the Rx. Thus, the distance between the $n$-th Tx antenna and the Rx is given by $r_n=\Vert \mathbf{r}-\mathbf{t}_n \Vert_2$, where $\mathbf{t}_n=[n_x \delta,n_y \delta,0]$ is the coordinate of the $n$-th antenna element. 

\begin{figure}[tbp]
	\centering
	\subfigure[\label{wavenumber_pattern}]{
		\includegraphics[height=3.46cm]{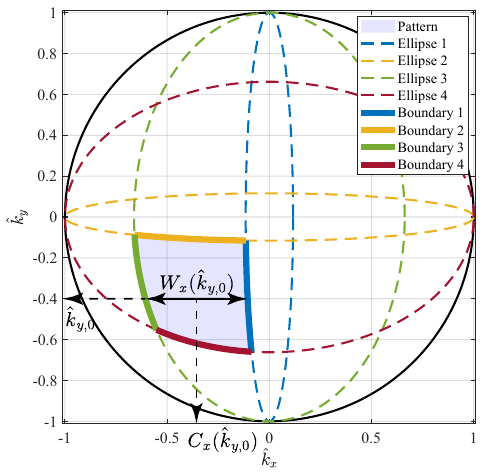}
	}
	\hspace{-5mm}
	\subfigure[\label{projection}]{
		\includegraphics[height=3.46cm]{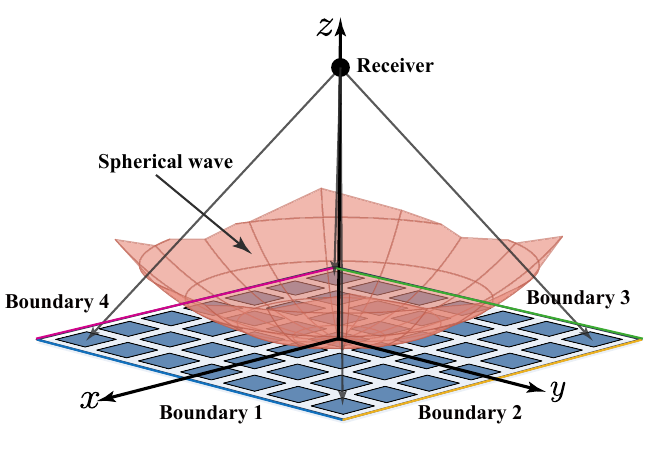}
	}
	\caption{Relationship between the wavenumber-domain pattern and the UPA geometry. (a) Wavenumber-domain pattern and its four boundaries. (b) Four boundaries of the UPA, where each boundary is determined by the receiver position and one edge of the array. The number of antenna elements are $N_x=N_y=511$, the central carrier frequency is $f_{\mathrm c}=100$ GHz, and the receiver location is $\mathbf{r}=[0.5,0.5,1]^T$.}
	\label{pattern_boundary}
\end{figure}

To provide a unified channel representation across both near-field and far-field regions, the FPW representation \cite{pizzo-1} is adopted in this paper, which in essence employs a series of orthogonal basis functions to model the spatial channel responses. Consequently, the channel response between the transceiver at subcarrier $m$ in \eqref{Eq1_1} can be rewritten as
\begin{equation}
\label{Eq1_2}
\mathbf{h}_{m} = \mathbf{h}_{m}^\text{w} \mathbf{\Psi}_{m}^H,
\end{equation}
where $\mathbf{h}_{m}^{\text{w}}\in\mathbb{C}^{1\times \vert\xi_m\vert}$ is the wavenumber-domain pattern at subcarrier $m$, and $\mathbf{\Psi}_{m}\in\mathbb{C}^{N\times\vert\xi_m\vert}$ is the wavenumber-domain dictionary matrix composed of Fourier plane-wave vectors. Let $\left(l_x,l_y\right)\in\xi_m$ denote the wavenumber indices associated with the $x$- and $y$-axes, and define $l \triangleq \left(l_x,l_y\right)\in \xi_m$. The $\left(n,l\right)$-th entry of the dictionary matrix $\mathbf{\Psi}_m$ in \eqref{Eq1_2} is given by \cite{pizzo-1}
\begin{equation}
\label{Eq1_3}
\begin{aligned}
\left[\mathbf{\Psi}_m\right]_{n,l} &= \frac{1}{\sqrt{N}}\exp{\left\{-j\left(\frac{2\pi l_x n_x}{N_x} + \frac{2\pi l_y n_y}{N_y}\right)\right\}}\\ 
& =\frac{1}{\sqrt{N}}\exp\left\{-jk_m \left(\hat{k}_{m,x}\delta n_x+\hat{k}_{m,y}\delta n_y\right)\right\},
\end{aligned}
\end{equation}
where $\hat{k}_{m,x}\triangleq\frac{2l_x}{\eta_m N_x}=\sin\theta \cos \phi$ and $\hat{k}_{m,y}\triangleq\frac{2l_y}{\eta_m N_y}=\sin\theta \sin\phi$ denote the normalized wavenumber along the $x$- and $y$-axes, with $\eta_m\triangleq f_m/f_\text{c}$. 
For each subcarrier, $\mathbf{\Psi}_m$ comprises a set of orthogonal discrete basis functions, namely the columns of a 2D DFT matrix. These basis functions are fixed and independent of the scattering environment. Importantly, their forms remain applicable to both narrowband and wideband systems. The distinction between these two systems lies in the wavenumber-domain support $\xi_m$, which collects all the wavenumber indices associated with subcarrier $m$. For a narrowband system operating at frequency $f_m$ with antenna spacing $\lambda_m / 2$, the support $\xi_m$ can be characterized by~\cite{pizzo-1}
\begin{equation}
\label{Eq1_4}
\xi_m = \left\{\left(l_x,l_y\right)\in\mathbb{Z}^2 \,\big|\, \left(\frac{\lambda_m l_x}{L_x}\right)^2+\left(\frac{\lambda_m l_y}{L_y}\right)^2 \le 1 \right\},
\end{equation}
where $\vert\xi_m\vert$ is the cardinality of $\xi_m$. As depicted in Fig. \ref{wavenumber_pattern}, the normalized wavenumbers collected by $\xi_m$ correspond exclusively to the propagating wave components confined within the black circle. Conversely, the evanescent wave components lying outside this circle are disregarded in the scope of this paper. The extension of this support to wideband systems will be proposed and discussed in Section \uppercase\expandafter{\romannumeral3}.

%\begin{figure}[tbp]
%    \centering
%    \includegraphics[width=7cm]{Figs/wavenumber_pattern.pdf}
%    \caption{Four boundaries of the wavenumber domain pattern. The number of antenna elements are $N_x=N_y=511$ and the central carrier frequency is $f_{\text{c}}=100$ GHz. For illustrative purposes, $\mathbf{r}=[0.5,0.5,1]^T$.} 
%    \label{wavenumber_pattern}
%\end{figure}

\subsection{Boundaries of the Wavenumber-Domain Pattern}
In our previous work\cite{ellipse}, we demonstrated that the boundaries of the wavenumber-domain pattern $\mathbf{h}_m^{\text{w}}$ for a UPA are characterized by four distinct semi-ellipses, as illustrated in Fig. \ref{wavenumber_pattern}. More importantly, these boundaries are jointly determined by the receiver's spatial location and the four physical edges of the UPA aperture. To provide a clearer geometric interpretation, \textbf{boundary 1} is taken as an example to illustrate the derivation of its analytical expression, denoted by $\mathcal{B}_{1}$. As shown in Fig. \ref{projection}, there is a direct mapping between the spatial geometry and the wavenumber-domain pattern. Specifically, each point on \textbf{boundary 1} in the wavenumber domain corresponds to the propagation direction originating from the receiver location $\mathbf{r}$ and pointing toward the array edge defined by $[L_x/2, y, 0]^T$ with $y \in [-L_y/2, L_y/2]$. Let $\mathbf{\hat{k}}=[\hat{k}_{x},\hat{k}_y,\hat{k}_z]^T$ be the normalized wavenumber vector, where $\hat{k}_z=\sqrt{1-\hat{k}_x^2-\hat{k}_y^2}$. Every point $(\hat{k}_x,\hat{k}_y)$ on $\mathcal{B}_{1}$ satisfies the directional constraint imposed by this geometric relationship, which is expressed as
\begin{equation}
\label{Eq1_5}
\mathbf{\hat{k}}=\text{dir} \left(\mathbf{r} - [L_x/2, y, 0]^T \right), \forall(\hat{k}_x,\hat{k}_y) \in \mathcal{B}_1,
\end{equation}
where $\text{dir}(\mathbf{a})=\mathbf{a}/\text{norm}(\mathbf{a})$ is the normalization operator. By expanding \eqref{Eq1_5}, the components $\hat{k}_x$ and $\hat{k}_y$ on $\mathbf{\mathcal{B}}_1$ are explicitly expressed as
\begin{equation}
\label{Eq1_6}
\hat{k}_x = (x_r - L_x/2) / \text{norm}\left([x_r-L_x/2,y_r-y,z_r]^T \right),
\end{equation}
\begin{equation}
\label{Eq1_7}
\hat{k}_y = (y_r - y) / \text{norm}\left([x_r-L_x/2,y_r-y,z_r]^T \right).
\end{equation}

%\begin{figure}[tbp]
%    \centering
%    \includegraphics[width=7cm]{Figs/projection_v2.pdf}
%    \caption{\textcolor{blue}{Four boundaries of the UPA. The boundaries of the wavenumber domain pattern are determined by the receiver's position and the array edges.}}
%    \label{projection}
%\end{figure}

Dividing $\hat{k}_y$ in \eqref{Eq1_7} by $\hat{k}_x$ in \eqref{Eq1_6} yields an expression for the spatial coordinate $y$ as
\begin{equation}
\label{Eq1_8}
y=\frac{\hat{k}_y}{\hat{k}_x}(L_x/2-x_r) + y_r, -L_y / 2 < y < L_y/2.
\end{equation}

By removing the finite extent constraint (i.e., $-L_y/2<y<L_y/2$) and substituting \eqref{Eq1_8} back into \eqref{Eq1_6} and \eqref{Eq1_7}, we can derive the closed-form expression for the semi-ellipse $\mathbf{\mathcal{E}}_1$ shown in Fig. \ref{wavenumber_pattern}, given by
\begin{equation} 
\label{Eq1_9}
\begin{aligned}
\mathcal{E}_{1} \triangleq \big\{ (\hat{k}_x, \hat{k}_y) &\in \mathbb{R}^2 \,\big|\,  C_1 \hat{k}_x^2 + \hat{k}_y^2 = 1, \\
& \text{sign}(\hat{k}_x) = \text{sign}(x_r - L_x/2)
\big\},
\end{aligned}
\end{equation}
where $\text{sign}(\cdot)$ is the sign function and $C_1=\frac{(L_x/2-x_r)^2+z_r^2}{(L_x/2-x_r)^2}$. Here, the geometric relationship between the receiver and the Tx array determines which specific half of the ellipse is valid, defined by the constraint $\text{sign}(\hat{k}_x) = \text{sign}(x_r - L_x/2)$.

Following the same procedure, the expressions for the remaining three semi-ellipses are derived as \cite{ellipse}
\begin{equation} 
\label{Eq1_10}
\begin{aligned}
\mathcal{E}_{2} \triangleq \big\{  (\hat{k}_x, \hat{k}_y) &\in \mathbb{R}^2 \,\big|\, C_2 \hat{k}_y^2 + \hat{k}_x^2 = 1 , \\
& \text{sign}(\hat{k}_y) = \text{sign}(y_r - L_y/2) \big\},
\end{aligned}
\end{equation}
\begin{equation} 
\label{Eq1_11}
\begin{aligned}
\mathcal{E}_{3} \triangleq \big\{  (\hat{k}_x, \hat{k}_y) & \in \mathbb{R}^2 \,\big|\, C_3 \hat{k}_x^2 + \hat{k}_y^2 = 1, \\
& \text{sign}(\hat{k}_x) = \text{sign}(x_r + L_x/2) \big\},
\end{aligned}
\end{equation}
\begin{equation} 
\label{Eq1_12}
\begin{aligned}
\mathcal{E}_{4} \triangleq \big\{  (\hat{k}_x, \hat{k}_y) &\in \mathbb{R}^2 \,\big|\, C_4 \hat{k}_y^2 + \hat{k}_x^2 = 1 , \\
& \text{sign}(\hat{k}_y) = \text{sign}(y_r + L_y/2) \big\},
\end{aligned}
\end{equation}
where $C_2=\frac{(L_y/2-y_r)^2+z_r^2}{(L_y/2-y_r)^2}$, $C_3=\frac{(L_x/2+x_r)^2+z_r^2}{(L_x/2+x_r)^2}$, and $C_4=\frac{(L_y/2+y_r)^2+z_r^2}{(L_y/2+y_r)^2}$. Building upon these semi-elliptical boundaries, Section \uppercase\expandafter{\romannumeral4} extracts two 1D scalar features, namely the pattern width and center, as shown in Fig.~\ref{wavenumber_pattern}. Based on these features, we derive exact closed-form expressions that establish a rigorous mapping to the receiver's spatial coordinates, enabling rapid, accurate, and robust beam training.

\section{Wavenumber domain pattern zooming}
This section first presents the derivation of the FPW representation for the wideband channel. Subsequently, we introduce and analyze the wavenumber domain pattern zooming effect, which inherently emerges when plane-wave transmission is employed across both far-field and near-field regions in wideband systems.

\subsection{FPW Representation of the Wideband Channel}
To extend the narrowband wavenumber-domain support defined in \eqref{Eq1_4} to wideband systems, we first consider an intuitive generalization by setting the antenna spacing to $\delta=\lambda_c / 2$, with $L_x=N_x\delta$ and $L_y=N_y \delta$, where $f_c$ denotes the central carrier frequency as specified in Section \uppercase\expandafter{\romannumeral2}-A. This configuration leads to the following frequency-dependent representation
\begin{equation}
\label{Eq2_4}
\xi_m = \left\{\left(l_x,l_y\right)\in\mathbb{Z}^2 \big| \left(\frac{2 l_x}{\eta_m N_x}\right)^2+\left(\frac{2 l_y}{\eta_m N_y}\right)^2 \le 1 \right\},
\end{equation}
which implies that $\vert \xi_m \vert$ grows proportionally with the subcarrier frequency $f_m$. Specifically, the wavenumber-domain support at higher frequencies (i.e., $f_m > f_{\text{c}}$), denoted as $\xi_\text{H}$, encompasses a greater number of wavenumber indices than that at the center frequency $f_{\text{c}}$.\footnote{Throughout this paper, the subscripts $\text{H}$, $\text{c}$, and $\text{L}$ denote subcarriers with frequencies higher than, equal to, and lower than the center frequency, respectively.} Meanwhile, the normalized wavenumber resolution between adjacent indices, defined by $\frac{2\pi/L_x}{k_m}=\lambda_m/L_x=\frac{2}{N_x \eta_m}$, becomes increasingly fine as the frequency increases. Conversely, for lower frequencies ($f_m < f_\text{c}$), the support contains fewer wavenumber indices and exhibits larger normalized wavenumber resolution. Notably, \eqref{Eq2_4} provides an intuitive extension of the narrowband model, and both this expression and the conclusions derived from it will be rigorously examined in the following analysis. 

Building upon the preceding observations, we next examine the structure and properties of the wavenumber-domain dictionary in \eqref{Eq1_2} across different subcarriers. Crucially, the frequency-dependent variation in the wavenumber-domain support described in \eqref{Eq2_4} directly manifests in the structure of the dictionary matrix $\mathbf{\Psi}_m$ in \eqref{Eq1_3}. Conversely, this relationship implies that by exploiting the intrinsic properties of $\mathbf{\Psi}_m$, one can derive an accurate and physically consistent expression of the wavenumber-domain support, thereby obtaining a rigorous FPW representation of the wideband channel. Therefore, we proceed to discuss the dictionary matrix for different subcarriers in the following. For clarity of expression, we temporarily account for the evanescent wave components previously omitted in \eqref{Eq2_4}. Detailed discussions regarding the distinction between the evanescent and propagating wave components are available in~\cite{wd_codebook, Nyquist_pizzo}. Under this theoretical assumption, the wavenumber-domain support $\xi_{\text{c}}$ collects all indices corresponding to normalized wavenumbers within the physically feasible range $\mathcal{R}\triangleq[-1,1]\times[-1,1]$. Consequently, the dictionary matrix $\mathbf{\Psi}_\text{c}$ at the central carrier frequency $f_{\text{c}}$ coincides with a complete 2D DFT matrix. 

From \eqref{Eq2_4}, the number of available wavenumber indices at lower carrier frequencies is smaller than that at $f_c$. Consequently, the dictionary matrix $\mathbf{\Psi}_\text{L}$ reduces to a partial DFT matrix, comprising only a subset of the full DFT columns. This reduction is physically well-justified since the $N_x \times N_y$ antennas, acting as spatial sampling points, become overly dense at lower frequencies. Specifically, the fixed spacing $\delta=\lambda_c/2$ is less than the half-wavelength sampling interval $\lambda_{\text{L}}/2$ as the frequency decreases. As a result, the rectangular wavenumber domain synthesized by these $N_x \times N_y$ samples (represented by the red solid rectangle in Fig. \ref{zoom_low}) covers an area significantly larger than the visible wavenumber region $\mathcal{R}$. This visible region is bounded by the physically possible angles, namely the elevation $\theta$ and azimuth $\phi$. Consequently, normalized wavenumbers residing outside the unit circle in Fig. \ref{zoom_low} correspond to either physically non-existent components or evanescent waves.

Conversely, at higher carrier frequencies ($f_m > f_\text{c}$), the fixed antenna spacing $\delta=\lambda_\text{c}/2$ becomes relatively large with respect to the wavelength, such that $\delta>\lambda_{\text{H}}/2$. This condition leads to spatial undersampling. Under these circumstances, the dictionary matrix $\mathbf{\Psi}_{\text{H}}$, based on the wavenumber-domain support defined in \eqref{Eq2_4}, exhibits an aliased DFT structure. Physically, this occurs because the $N_x \times N_y$ antenna array provides an insufficient spatial sampling rate to resolve the fine angular components of $\mathbf{h}_{\text{H}}^{\text{w}}$ associated with high wavenumbers. As a consequence, these high-wavenumber components alias into low-wavenumber regions. Mathematically, these high-wavenumber contents fold into lower wavenumber indices due to the periodicity of the complex exponential terms in \eqref{Eq1_3} and the finite sampling grid. This aliasing phenomenon is clearly illustrated in Fig. \ref{zoom_high}, where energy from high-wavenumber regions maps onto lower-index locations of the DFT grid, thereby producing spurious low-wavenumber responses in $\mathbf{h}_\text{H}^{\text{w}}$.

Hence, by integrating the insights gained from the analysis of the central, high, and low-frequency dictionary matrices, the initial support $\xi_m$ defined in \eqref{Eq2_4} must be refined to rigorously characterize the wideband FPW representation as
\begin{equation}
\label{Eq2_5}
\begin{aligned}
\xi_m^{\prime}=\Bigg\{\left(l_x,l_y\right)\in\mathbb{Z}^2\,\big|\,\vert l_x\vert\le\frac{L_x}{2\delta},\vert l_y\vert\le\frac{L_y}{2\delta}, \Bigg. \\ \Bigg. \left(\frac{2 l_x}{N_x}\right)^2+\left(\frac{2 l_y}{N_y}\right)^2\le \left(\frac{f_m}{f_\text{c}}\right)^2 \Bigg\},
\end{aligned}
\end{equation}
which represents the intersection of the DFT rectangular region and the normalized wavenumber circle region. Specifically, the first pair of constraints on $l_x$ and $l_y$ effectively resolves the aliasing issue at high frequencies by confining the wavenumber indices within the fundamental DFT period. Meanwhile, the last constraint mitigates the redundancy at lower frequencies by retaining only those indices corresponding to physically existing propagating waves, namely those lying within the unit normalized wavenumber circle.

\subsection{Pattern Zooming Effect}
\begin{figure*}[htbp]
\centering
\subfigure[\label{zoom_low}]{ \includegraphics[width=6cm]{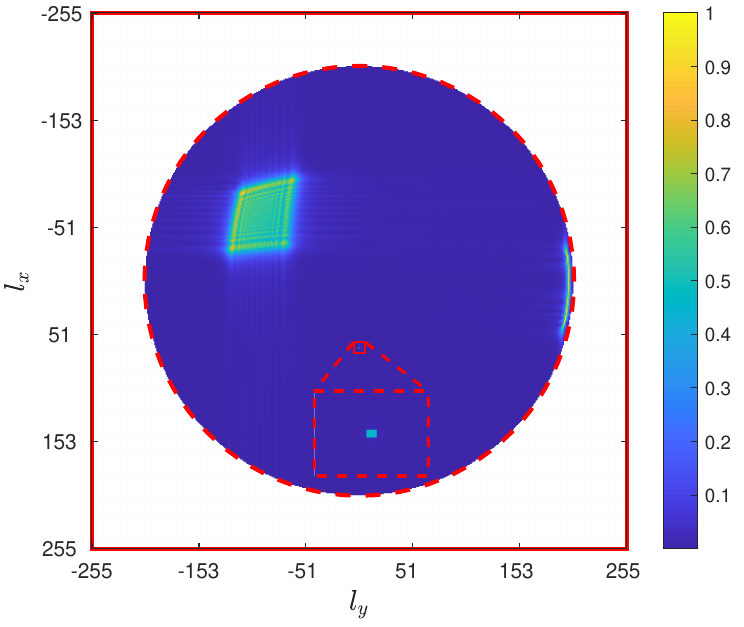}
}
\hspace{-6mm}
\subfigure[\label{zoom_center}]{
\includegraphics[width=6cm]{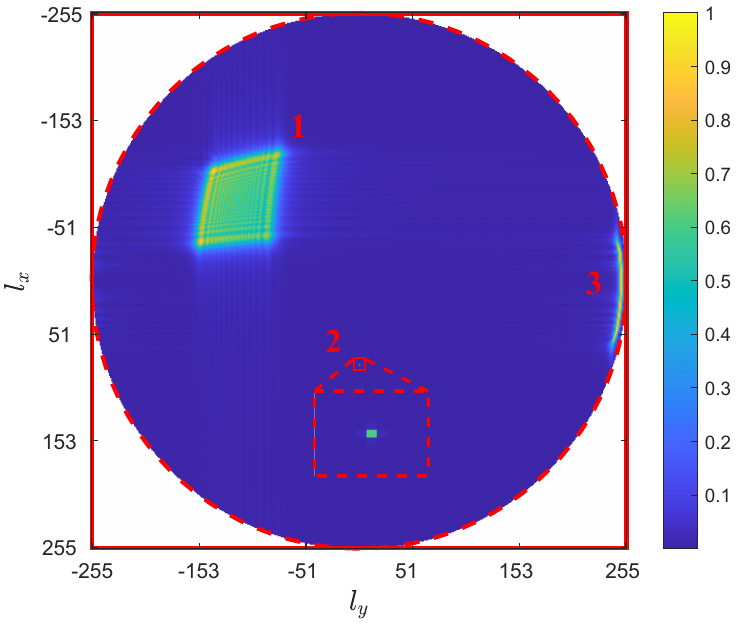}
}
\hspace{-6mm}
\subfigure[\label{zoom_high}]{
\includegraphics[width=6cm]{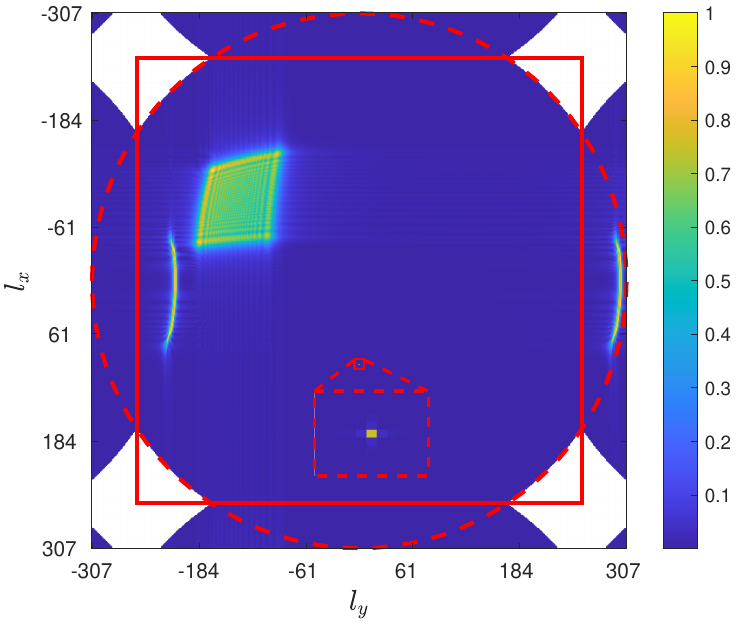}
}
\caption{The schematic diagram of the wavenumber domain pattern zooming in wideband systems. The number of antenna elements along the $x$-axis and $y$-axis are $N_x=N_y=511$. The central carrier frequency is $f_{\text{c}}=100$ GHz, and the bandwidth is $B=20$ GHz. The normalized wavenumber-domain patterns of three subcarriers are presented: (a) the lowest frequency  $f_\text{L}$, (b) the center frequency $f_{\text{c}}$, and (c) the highest frequency $f_{\text{H}}$. Channel (1) corresponds to a near-field receiver at $l_{x,(1)}=-20$, $l_{y,(1)}=-250$, and $r_{(1)}=2$; channel (2) to a far-field receiver at $l_{x,(2)}=-80$, $l_{y,(2)}=0$, and $r_{(2)}=800$; and channel (3) to another near-field receiver at $l_{x,(3)}=80$, $l_{y,(3)}=120$, and $r_{(3)}=3$.}
\label{pattern_zooming}
\end{figure*}

As established in Section \uppercase\expandafter{\romannumeral3}-A, the normalized wavenumber resolution varies across subcarriers in wideband systems. This frequency-dependent variation gives rise to a phenomenon we term the pattern zooming effect. To illustrate this phenomenon, the wavenumber-domain patterns of different subcarriers are presented. Specifically, since the dictionary matrix $\mathbf{\Psi}_m$ is a semi-unitary matrix, satisfying $\mathbf{\Psi}_m^H \mathbf{\Psi}_m = \mathbf{I}_{\vert \xi_m \vert}$, the wavenumber-domain pattern $\mathbf{h}_m^{\text{w}}$ can be derived from \eqref{Eq1_2} as
\begin{equation}
\label{Eq2_6}
\mathbf{h}_m^{\text{w}}=\mathbf{h}_m \mathbf{\Psi}_m,
\end{equation}
where the wavenumber-domain pattern can thus be interpreted as the gain obtained from a noise-free exhaustive sweeping using a PS-based wavenumber-domain codebook composed of plane-wave codewords\footnote{Throughout this paper, the terms codebook and dictionary matrix are used interchangeably, both referring to $\mathbf{\Psi}_m$}. Fig. \ref{pattern_zooming} compares the wavenumber-domain patterns at the lowest, center, and highest subcarrier frequencies. The red dashed circle marks the wavenumber-domain support $\xi_m$ defined in \eqref{Eq2_4}, while the red solid rectangle indicates the entire wavenumber region synthesized by performing a 2D DFT on the spatial channel $\mathbf{h}_m$. Three representative channel cases are illustrated in Fig. \ref{pattern_zooming}, with their corresponding positions marked by indices 1, 2, and 3 in Fig. \ref{zoom_center}.\footnote{For enhanced clarity of visualization, a relatively large bandwidth and a short near-field distance are adopted in these figures. This choice, however, does not compromise the generality of the analysis or the validity of the conclusions.} By examining the FPW representation in \eqref{Eq1_2} and Fig.~\ref{zoom_center}, one can observe that a key distinction: a far-field channel can be approximated by a single plane wave, whereas a near-field spherical-wave channel should be approximated by a linear superposition of multiple plane waves\cite{pizzo-1,wch-twc}.

Comparing the three patterns for \textbf{channel 1} in Fig. \ref{pattern_zooming}, it is evident that both the width and the center of the pattern, as measured by the wavenumber indices, vary significantly across subcarriers. This phenomenon, termed the pattern zooming effect, arises because the normalized wavenumber resolution differs among subcarriers, whereas the normalized wavenumber width and center of the receiver's pattern remain constant across subcarriers, which will be demonstrated in Section \uppercase\expandafter{\romannumeral4}. This effect is analogous to observing the pattern at the center frequency through a magnifying or reducing lens. For \textbf{channel 2}, where the normalized wavenumber width of the far-field pattern is negligible, the primary variation observed across subcarriers is manifested as a shift in the center of the wavenumber indices. For \textbf{channel 3}, which corresponds to a near-field receiver at a high wavenumber, wavenumber aliasing occurs, consistent with the analysis in Section \uppercase\expandafter{\romannumeral3}-A. 

It is crucial to emphasize that while the pattern zooming effect shares the frequency-dependent nature of wideband systems with the well-known beam split and rainbow effects, it represents a distinct phenomenon. Specifically, the beam split effect typically arises in conventional far-field wideband systems employing plane-wave transmission, such as traditional far-field beam training utilizing DFT codebooks, manifesting as a frequency-dependent angular deviation of the beam peak~\cite{far-hierarchical}. Conversely, the rainbow effect occurs in conventional near-field wideband systems utilizing spherical-wave transmission, such as near-field beam training with polar-domain codebooks, where the focused beam peak shifts across both angle and distance domains~\cite{narrow_near, near-hierarchical-1}. In contrast, the pattern zooming effect emerges when the recently emerging unified beam training paradigms, which utilize plane-wave transmission based on wavenumber-domain codebooks across both far-field and near-field regions~\cite{dft_bt_1,dft_bt_2,dft_bt_3,dft_bt_4}, are applied to wideband systems. Furthermore, these phenomena correspond to different beam training methodologies. Conventional beam training determines the optimal transmission beam by searching for the codeword with the maximum received gain. Consequently, the beam split and rainbow effects are primarily concerned with the frequency-dependent shift of the specific beam peak. However, under the unified beam training paradigm utilizing wavenumber-domain codebooks, the receiver parameters are determined through the mathematical analysis of the collected pattern rather than peak searching. Therefore, the pattern zooming effect uniquely manifests as the frequency-dependent scaling of the entire beam pattern, rather than a mere spatial shift of the peak gain.

\section{Pattern zooming based near-field wideband beam training with wavenumber-domain codebook}
In this section, we propose an efficient near-field wideband beam training method leveraging the pattern zooming effect. Inspired by~\cite{rainbow}, we adopt a frequency-dependent TD beamforming architecture at the Tx, where each antenna element is integrated with a dedicated true time delay (TTD) circuit, referred to as the Full-TTD architecture. Recent studies have shown that, compared with conventional Non-TTD and Sparse-TTD architectures employing PSs, the Full-TTD architecture does not incur significant penalties in hardware cost, optimization complexity, quantization, or implementation loss~\cite{Full-TTD}. Instead, it offers superior engineering viability for wideband and ultra-wideband systems. This architecture is employed for two reasons. First, it allows for controllable pattern zooming during the beam training stage, which substantially accelerates the process. Second, it facilitates the suppression of the detrimental effects of the beam squint effect during data transmission. Building upon this architecture, the proposed beam training method comprises two primary stages. The first stage focuses on acquiring the complete wavenumber-domain pattern, achieved through a specialized probing approach that utilizes controllable pattern zooming, as detailed in Section \uppercase\expandafter{\romannumeral4}-A. The second stage determines the receiver's spatial coordinates by extracting specific 1D scalar features, namely the pattern width and center, from the acquired pattern. Based on these extracted features, we derive exact closed-form expressions for accurate joint angle and distance estimation, which is presented in Section \uppercase\expandafter{\romannumeral4}-B. Finally, we evaluate the computational complexity of the proposed scheme in Section \uppercase\expandafter{\romannumeral4}-C, discuss its extension to multipath environments in Section \uppercase\expandafter{\romannumeral4}-D, and analyze the required maximum delay range in Section \uppercase\expandafter{\romannumeral4}-E.

\subsection{Pattern Zooming Based Wideband Beam Training}
The frequency-domain response of the $n$-th TD circuit at frequency $f_m$ is given by $\exp(-j2\pi f_m \tau_n^{\prime})$\cite{rainbow}, where $\tau_n^{\prime}\triangleq r_n^{\prime}/c$ denotes the adjustable delay of the $n$-th unit. During the beam training stage, we configure $r_n^{\prime}$ as
\begin{equation}
\label{Eq3_1}
r_n^{\prime} = n_x\delta\frac{2 l_x^{\prime}}{N_x} + n_y\delta \frac{2 l_y^{\prime}}{N_y},
\end{equation}
where $l_x^{\prime}$ and $l_y^{\prime}$ represent the adjustable TD parameters. Thus, the frequency-dependent wavenumber-domain codeword at $f_m$ can be expressed as
\begin{equation}
\label{Eq3_2}
\left[\mathbf{\psi}(l_x^{\prime},l_y^{\prime})\right]_n=\frac{1}{\sqrt{N}} \exp \left\{-jk_m\left(n_x\delta \frac{2 l_x^{\prime}}{N_x} + n_y \delta \frac{2 l_y^{\prime}}{N_y}\right) \right\}.
\end{equation}
Then, the beam gain produced by the TD beamforming on subcarrier $m$ can be written as 
\begin{equation}
\begin{aligned}
\label{Eq3_3}
G=\frac{1}{N} \left\vert D_{\tilde{N}_x}\left(\frac{2\pi(l_x-\eta_m l_{x}^{\prime})}{N_x}\right) D_{\tilde{N}_y}\left(\frac{2\pi(l_y- \eta_ml_y^{\prime})}{N_y}\right) \right \vert,
\end{aligned}
\end{equation}
where the function $D_N(x)=\frac{\sin((N+1/2)x)}{\sin(x/2)}$ is the Dirichlet kernel.
Because the Dirichlet kernel is periodic in $l_x$ and $l_y$ with periods $N_x$ and $N_y$, respectively, the aligned normalized wavenumbers at frequency $f_m$ can be expressed as
\begin{equation}
\label{Eq3_4}
\left(\hat{k}_{m,x},\hat{k}_{m,y}\right)=\left(\frac{2l_x^{\prime}}{N_x}+\frac{2 p_x}{\eta_m},\frac{2l_{y}^{\prime}}{N_y}+\frac{2 p_y}{\eta_m}\right),
\end{equation}
where $p_x,p_y\in\mathbb{Z}$ represent the periodicity indices. As established, \eqref{Eq3_4} elucidates why both this work and existing studies on wideband beam training favor the TD beamforming architecture. In particular, by configuring the TD parameters such that both $2l_x^{\prime}/N_x$ and $2l_y^{\prime}/N_y$ fall within the feasible range $\mathcal{R}$, we obtain $p_x=p_y=0$. This configuration ensures that all subcarriers align with the identical normalized wavenumber, thereby eliminating the detrimental effects of pattern zooming during data transmission. Conversely, by configuring the TD parameters such that at least one of these terms falls outside the feasible range $\mathcal{R}$, different subcarriers will align with distinct normalized wavenumbers. This enables the TD beamforming to intentionally induce the pattern zooming effect to probe multiple normalized wavenumbers simultaneously, thereby significantly accelerating the beam training process. For instance, when the configuration places $2l_x^{\prime}/N_x$ outside the feasible range $\mathcal{R}$ while keeping $2l_y^{\prime}/N_y$ within it, the different subcarriers will sweep across distinct values of $\hat{k}_x$ while maintaining a perfectly aligned identical $\hat{k}_y$. Furthermore, the extent of the pattern zooming effect is determined by the values of $l_x^{\prime}$ and $l_y^{\prime}$, which specify the corresponding integers $p_x$ and $p_y$. Larger absolute values of these periodicity indices result in a wider spread of the subcarrier probing directions.

\subsection{Wavenumber-Domain Pattern Based Joint Angle and Distance Accurate Estimation}
As analyzed in Section \uppercase\expandafter{\romannumeral2}-C, determining the receiver coordinates based on the entire boundary of the wavenumber-domain pattern requires the perfect extraction of complete 2D semi-ellipses. However, relying on such complete boundary extraction is stringent. In certain practical scenarios, such as in multipath environments, obtaining complete boundaries becomes highly challenging due to overlapping patterns, which renders ellipse fitting approaches ineffective. To resolve this issue, we propose a robust feature extraction strategy that reduces the required dimensionality. Rather than depending on the complete 2D boundaries, we reveal that the spatial coordinates can be accurately obtained using specific 1D scalar features, namely the pattern width and center. These scalar features remain obtainable even when multipath components induce partial pattern overlaps, with a detailed discussion on this multipath extension deferred to Section \uppercase\expandafter{\romannumeral4}-D.

As illustrated in Fig. \ref{wavenumber_pattern}, we define $W_x({\hat{k}_{y,0}})$ and $C_x({\hat{k}_{y,0}})$ as the normalized wavenumber width and center along the $\hat{k}_x$-axis, respectively, for a fixed normalized wavenumber $\hat{k}_y=\hat{k}_{y,0}$. Similarly, $W_y({\hat{k}_{x,0}})$ and $C_y({\hat{k}_{x,0}})$ are defined as the normalized wavenumber width and center along the $\hat{k}_y$-axis for a fixed $\hat{k}_x=\hat{k}_{x,0}$. Their expressions can be derived from the boundary expressions in \eqref{Eq1_9}-\eqref{Eq1_12} and are given as
\begin{equation}
\label{Eq3_5}
W_x(\hat{k}_{y,0}) =
\begin{cases}
K_{y,0}\left \vert\frac{1}{\sqrt{C_3}}-\frac{1}{\sqrt{C_1}}\right\vert,\vert x_r\vert >  L_x/2,\\
K_{y,0}\left(\frac{1}{\sqrt{C_3}}+\frac{1}{\sqrt{C_1}}\right),\vert x_r\vert \le  L_x/2, \\
\end{cases}
\end{equation}
\begin{equation}
\label{Eq3_6}
C_x(\hat{k}_{y,0}) =
\begin{cases}
-\frac{K_{y,0}}{2}\left(\frac{1}{\sqrt{C_3}}+\frac{1}{\sqrt{C_1}}\right),x_r<-L_x/2,\\
\frac{K_{y,0}}{2}\left(\frac{1}{\sqrt{C_3}}-\frac{1}{\sqrt{C_1}}\right),\vert x_r\vert \le  L_x/2, \\
\frac{K_{y,0}}{2}\left(\frac{1}{\sqrt{C_3}}+\frac{1}{\sqrt{C_1}}\right),x_r>L_x/2, \\
\end{cases}
\end{equation}
\begin{equation}
\label{Eq3_7}
W_y(\hat{k}_{x,0}) = 
\begin{cases}
K_{x,0}\left \vert \frac{1}{\sqrt{C_4}}-\frac{1}{\sqrt{C_2}}\right \vert,\vert y_r\vert >  L_y/2,\\
K_{x,0}\left(\frac{1}{\sqrt{C_4}}+\frac{1}{\sqrt{C_2}}\right),\vert y_r\vert \le  L_y/2, \\
\end{cases}
\end{equation}
\begin{equation}
\label{Eq3_8}
C_y(\hat{k}_{x,0}) =
\begin{cases}
-\frac{K_{x,0}}{2}\left(\frac{1}{\sqrt{C_4}}+\frac{1}{\sqrt{C_2}}\right),y_r<-L_y/2,\\
\frac{K_{x,0}}{2}\left(\frac{1}{\sqrt{C_4}}-\frac{1}{\sqrt{C_2}}\right),\vert y_r\vert \le  L_y/2, \\
\frac{K_{x,0}}{2}\left(\frac{1}{\sqrt{C_4}}+\frac{1}{\sqrt{C_2}}\right),y_r>L_y/2, \\
\end{cases}
\end{equation}
where the auxiliary parameters are defined as $K_{y,0}=\sqrt{1-\hat{k}_{y,0}^2}$ and $K_{x,0}=\sqrt{1-\hat{k}_{x,0}^2}$. Given that $L_x$ and $L_y$ are fixed, these four quantities are piecewise functions determined exclusively by the geometric relationship between the array and the Rx. The specific branch taken by each quantity is specified by the position of the wavenumber-domain boundary. For instance, if \textbf{boundary 1} and \textbf{boundary 3} both lie on the negative $\hat{k}_x$-axis, this condition implies that $\text{sign}(x_r-L_x/2)=\text{sign}(x_r+L_x/2)=-1$, and therefore $x_r < -L_x/2$. 

Based on the definitions of $C_1$ and $C_3$ in Section \uppercase\expandafter{\romannumeral2}-C, the terms $1/\sqrt{C_1}$ and $1/\sqrt{C_3}$ can be expressed as $1/\sqrt{C_1}=\frac{L_x/2-x_r}{\sqrt{(L_x/2-x_r)^2+z_r^2}}$ and $1/\sqrt{C_3}=\frac{L_x/2+x_r}{\sqrt{(L_x/2+x_r)^2+z_r^2}}$, which represent the normalized wavenumbers from the receiver to the left and right edges of the array, respectively. Consequently, the normalized wavenumber width and center can be interpreted as the difference and the mean of the normalized wavenumbers spanning the array edges. Crucially, as the receiver moves relative to the array, the wavenumbers perceived across the points of the array vary accordingly. This implies that both the normalized wavenumber width and center are inherently position-dependent. Next, we detail how to predict the receiver's angle and distance by utilizing our exact expressions in \eqref{Eq3_5}-\eqref{Eq3_8} together with the measured wavenumber-domain pattern from Section \uppercase\expandafter{\romannumeral4}-A.

Without loss of generality, we assume $\vert x_r \vert < L_x /2$ and $\vert y_r \vert < L_y /2$. Based on the expressions in \eqref{Eq3_5} and \eqref{Eq3_6}, the auxiliary parameters $C_1$ and $C_3$ can be obtained as functions of the measured normalized wavenumber width $W_x(\hat{k}_{y,0})$ and center $C_x(\hat{k}_{y,0})$, as
\begin{equation}
\label{Eq3_9}
C_1 = \left(\frac{2K_{y,0}}{W_x(\hat{k}_{y,0}) - 2C_x(\hat{k}_{y,0})} \right)^2,
\end{equation}
\begin{equation}
\label{Eq3_10}
C_3 = \left(\frac{2K_{y,0}}{W_x(\hat{k}_{y,0}) + 2C_x(\hat{k}_{y,0})} \right)^2.
\end{equation}
By recalling the definitions of $C_1$ and $C_3$, and expressing $z_r^2$ in terms of $C_1$ and $C_3$ separately, we have
\begin{equation}
\label{Eq3_11}
(C_1-1)(L_x/2-x_r)^2 = z_r^2 = (C_3-1)(L_x/2+x_r)^2,
\end{equation}
which is a quadratic equation in $x_r$ and yields the following closed-form solution
\begin{equation}
\label{Eq3_12}
    x_r= \frac{L_x}{2} \left( \frac{\sqrt{C_1-1} -\sqrt{C_3-1}}{\sqrt{C_1-1} +\sqrt{C_3-1}} \right)^{2\mathbb{I}(\vert x_r \vert \le L_x /2)-1},
\end{equation}
where $\mathbb{I}(\cdot)$ is the indicator function. Notably, \eqref{Eq3_12} depends on the geometric relationship between the Rx and the array boundary, which, as discussed previously, can be determined from the location of the wavenumber-domain boundary.

Then, $y_r$ can be obtained in the same way as
\begin{equation}
\label{Eq3_13}
    y_r= \frac{L_y}{2} \left( \frac{\sqrt{C_2-1} -\sqrt{C_4-1}}{\sqrt{C_2-1} +\sqrt{C_4-1}} \right)^{2\mathbb{I}(\vert y_r \vert \le L_y /2)-1},
\end{equation}
where
\begin{equation}
\label{Eq3_14}
C_2 = \left(\frac{2K_{x,0}}{W_y(\hat{k}_{x,0}) - 2C_y(\hat{k}_{x,0})} \right)^2,
\end{equation}
\begin{equation}
\label{Eq3_15}
C_4 = \left(\frac{2K_{x,0}}{W_y(\hat{k}_{x,0}) + 2C_y(\hat{k}_{x,0})} \right)^2,
\end{equation}
which are derived under the assumption that $\vert y_r \vert < L_y /2$. Finally, $z_r$ can be derived by \eqref{Eq3_11} as
\begin{equation}
\label{Eq3_16}
z_r=\sqrt{(C_1-1)}\left \vert L_x/2-x_r \right\vert.
\end{equation}

\begin{algorithm}[t]
	\renewcommand{\algorithmicrequire}{\textbf{Require:}}
	\renewcommand{\algorithmicensure}{\textbf{Ensure:}}
	\caption{Pattern Zooming based Near-Field Wideband Beam Training with Wavenumber-Domain Codebook}
	\label{alg1}
	\begin{algorithmic}[1]
            \REQUIRE $L_x$, $L_y$, $f_c$, $B$.
            \ENSURE $(x_r,y_r,z_r)$.

            \STATE $f_{\text{L}}=f_c - \frac{B}{2}$, $f_{\text{H}}=f_c + \frac{B}{2}$. 
            
            \FOR{$l_y$ in $\{{-\tilde{N}_y},-\tilde{N}_y+1,\ldots,\tilde{N}_y\}$}
                \STATE Compute the potential range $[l_{x,\text{min}},l_{x,\text{max}}]=[-\frac{N_x}{2}-\frac{N_xl_y}{N_y},\frac{N_x}{2}-\frac{N_xl_y}{N_y}]$
                \STATE Compute the TD parameter $l_y^{\prime}=l_y$, $l_x^{\prime}=-N_x\lceil\max \left\{2f_{\text{L}}l_{x,\text{max}}/N_x,-2f_{\text{H}}l_{x,\text{min}}/N_x \right\} / B\rceil$
                \STATE Obtain the wavenumber-domain codeword $\mathbf{\psi}(l_x^{\prime},l_y^{\prime})$ using \eqref{Eq3_2}
                \STATE Compute received signal $y_m=\sqrt{P}\mathbf{h}_m \mathbf{\psi}(l_x^{\prime},l_y^{\prime}) + n_m$
                \STATE $\mathbf{h}_\text{c}^{\text{w}}\left(\frac{2l_x^{\prime}}{N_x}+\frac{2 p_x}{\eta_m},\frac{2l_{y}^{\prime}}{N_y}\right) \approx y_m$ according to \eqref{Eq3_4} 
            \ENDFOR
            
            \STATE Obtain $\vert \mathbf{h}_{\text{c},\text{filt}}^{\text{w}} \vert$ by filtering $\vert \mathbf{h}_{\text{c}}^{\text{w}} \vert$ with $\tau$
            \STATE Select the non-zero indices to obtain $\hat{k}_{x,0}$, $\hat{k}_{y,0}$ and estimate $W_x(\hat{k}_{y,0})$, $C_x(\hat{k}_{y,0})$, $W_y(\hat{k}_{x,0})$, and $C_y(\hat{k}_{x,0})$
            \STATE Compute $C_1$, $C_2$, $C_3$, $C_4$ using \eqref{Eq3_9}, \eqref{Eq3_14}, \eqref{Eq3_10}, and \eqref{Eq3_15} 
            \STATE Compute $x_r$, $y_r$, and $z_r$ using \eqref{Eq3_12}, \eqref{Eq3_13}, and \eqref{Eq3_16}
            \STATE \textbf{return} $x_r$, $y_r$, and $z_r$
	\end{algorithmic}  
\end{algorithm}
The specific procedure of the proposed pattern zooming based near-field wideband beam training with the wavenumber-domain codebook scheme is detailed in Algorithm \ref{alg1}. The initial operations from step 1 to step 8 in Algorithm \ref{alg1} focus on configuring the TD parameters $l_x^{\prime}$ and $l_y^{\prime}$ to induce controllable pattern zooming. Without loss of generality, we assume the codeword is aligned with the normalized wavenumber $\hat{k}_{\text{c},x} = 0 = \frac{2l_x^{\prime}}{N_x} + 2p_x$ at the center frequency, and configure the parameters such that $2l_x^{\prime}/N_x < -1$ and $\left|2l_y^{\prime}/N_y\right| < 1$. This specific configuration allows different subcarriers to simultaneously probe distinct $\hat{k}_{x}$ for a fixed $\hat{k}_y$, thereby significantly accelerating the acquisition of the complete wavenumber-domain pattern. Under this setup, we obtain $\frac{2l_x^{\prime}}{N_x}=-2p_x$, meaning $p_x=-\frac{l_x^{\prime}}{N_x}$ becomes a positive integer. For an arbitrary $l_y^{\prime}=l_y$, the entire potential range of $\hat{k}_x$ is bounded by $\left[-\sqrt{1-(2l_y^{\prime} / N_y)^2},\sqrt{1-(2l_y^{\prime} / N_y)^2} \right]$. The corresponding range in wavenumber indices is given by $[l_{x,\text{min}},l_{x,\text{max}}]= \frac{N_x}{2}\left[-\sqrt{1-(2l_y^{\prime} / N_y)^2},\sqrt{1-(2l_y^{\prime} / N_y)^2}\right]$. To ensure that the pattern zooming covers this entire potential wavenumber range for all subcarriers, the following constraints must be satisfied
\begin{equation}
\label{Eq3_17}
\frac{2 l_{x,\text{min}}}{N_x} \ge \frac{2l_x^{\prime}}{N_x} + \frac{2p_x f_{\text{c}}}{f_{\text{H}}},
\end{equation}
\begin{equation}
\label{Eq3_18}
\frac{2 l_{x,\text{max}}}{N_x} \le \frac{2l_x^{\prime}}{N_x} + \frac{2p_x f_{\text{c}}}{f_{\text{L}}}.
\end{equation}
One valid solution for $l_x^{\prime}$ that guarantees full coverage is
\begin{equation}
\label{Eq3_19}
l_x^{\prime}=-N_x\lceil\max \left\{2f_{\text{L}}l_{x,\text{max}}/N_x,-2f_{\text{H}}l_{x,\text{min}}/N_x \right\} / B\rceil,
\end{equation}
where $\lceil x\rceil$ is the ceiling function. Based on the mechanism of controllable pattern zooming in \eqref{Eq3_4}, we have $\mathbf{h}_\text{c}^{\text{w}}\left(\frac{2l_x^{\prime}}{N_x}+\frac{2 p_x}{\eta_m},\frac{2l_{y}^{\prime}}{N_y}\right) \approx y_m$ in step 7. Then, to alleviate the influence of noise, a filtering process is performed on the obtained pattern $\mathbf{h}_m$ in step 9. The filtered wavenumber-domain pattern is given by
\begin{equation}
\label{Eq3_20}
\vert \mathbf{h}_{\text{c},\text{filt}}^{\text{w}}(l_x,l_y)\vert = 
\begin{cases}
1, \vert \mathbf{h}_{\text{c}}^{\text{w}}(l_x,l_y) \vert \ge \tau \max (\vert \mathbf{h}_{\text{c}}^{\text{w}}\vert),\\
0, \vert \mathbf{h}_{\text{c}}^{\text{w}}(l_x,l_y) \vert < \tau \max (\vert \mathbf{h}_{\text{c}}^{\text{w}}\vert),\\
\end{cases}
\end{equation}
where $\tau$ denotes the threshold coefficient. Consistent with existing DFT codebook-based near-field beam training schemes~\cite{dft_bt_1, dft_bt_3, dft_bt_5}, the proposed algorithm adopts a relative thresholding strategy, where the threshold is defined as a fixed proportion of the maximum magnitude of the acquired wavenumber-domain pattern rather than a fixed absolute threshold. Since the peak magnitude of the acquired pattern varies with the channel conditions, the actual threshold automatically scales with the received pattern magnitude, thereby improving the robustness of support extraction. Finally, in steps 10-12, the filtered wavenumber-domain pattern is utilized to derive the receiver's angle and distance, enabling the Tx to generate the focusing beam for data transmission.

\subsection{Computational Complexity Analysis}
For the proposed method, the primary computational burden arises from the threshold filtering of the wavenumber-domain pattern, which incurs a complexity of $\mathcal{O}(N)$. After filtering, the extraction of the pattern width and center, as well as the subsequent coordinate calculation using the derived closed-form expressions, involve only a constant number of arithmetic operations, yielding a complexity of $\mathcal{O}(1)$. Therefore, the overall computational complexity of the proposed method is $\mathcal{O}(N)$. In contrast, conventional far-field and near-field beam training schemes require exhaustive gain comparisons over all codewords, resulting in computational complexities of $\mathcal{O}(N)$ and $\mathcal{O}(NS)$ respectively, where $S$ denotes the number of distance samples in the near-field codebook. Among existing pattern-based methods, the scheme in~\cite{dft_bt_2} exhibits a similar complexity of $\mathcal{O}(N)$ but relies on mathematical approximations. In contrast, the method in~\cite{dft_bt_1} requires an exhaustive search over $|\mathcal{Z}|$ distance grids, resulting in a computational complexity of $\mathcal{O}(|\mathcal{Z}|N)$.

\subsection{Discussion on Multipath Scenarios}
In practical mmWave and THz systems, non-line-of-sight (NLoS) paths typically suffer much larger path loss than the LoS path due to the weak reflection and diffraction capabilities of high-frequency electromagnetic waves. As a result, the LoS component usually dominates the channel, while the residual multipath components are sufficiently weak to be treated as background noise. Therefore, focusing on the dominant path is a widely adopted assumption, and these weak NLoS components do not significantly affect the proposed beam training procedure. In rich scattering environments where multipath propagation is intentionally exploited, beam training is commonly formulated as a channel estimation problem, followed by maximum ratio transmission based on the estimated channel~\cite{multipath}. Detailed wavenumber-domain channel estimation methods are provided in our prior work~\cite{wd_channel}. In scenarios where a strong NLoS path causes pattern overlapping with the dominant path, the proposed method remains robust. Instead of relying on complete pattern boundaries, it exploits the width and center features of the beam pattern. By measuring the pattern width along the orthogonal direction at different wavenumber coordinates, an abnormal increase in the width sequence can be observed within the overlapping interval. These anomalous regions can be identified and removed using algorithms like the multi-Otsu method~\cite{multi-Otsu}. Consequently, the width and center features can still be extracted from the remaining regions, enabling parameter estimation under multipath scenarios. The effectiveness of this mechanism is verified in Fig.~\ref{UPA_multipath}, where the proposed method accurately estimates the user's position despite partial overlap between the LoS and NLoS wavenumber-domain patterns.

\subsection{Maximum Delay Range Analysis}
This subsection analyzes the maximum delay range required by the TTD circuits, since the hardware cost of TTDs generally scales with the maximum delay range~\cite{rainbow}. The time delay of the $n$-th antenna is given by $\tau_n^{\prime}=n\frac{\delta\theta^{\prime}}{c}$, where $\theta^{\prime}=\frac{2l^{\prime}}{N}$. To ensure non-negative delays, a common delay offset is introduced, yielding the realistic delay $\bar{\tau}_n^{\prime}=\tau_n^{\prime}-\min_n\tau_n^{\prime}$. Since an abnormal wavenumber parameter $l^{\prime} < 0$ is used to induce the pattern zooming effect, $\tau^{\prime}_n$ or $\bar{\tau}_n^{\prime}$ decreases monotonically with respect to $n$, and the maximum delay range is therefore given by $\Delta\tau_{\text{plane-wave}}=\frac{4\delta|l^{\prime}|}{c}$. According to~\eqref{Eq3_19}, a larger system bandwidth requires a smaller $|l^{\prime}|$ to cover the same wavenumber range. Since $\Delta\tau_{\text{plane-wave}}$ is proportional to $|l^{\prime}|$, increasing the bandwidth directly reduces the required delay range. For comparison, conventional near-field wideband beam training relies on spherical-wave focusing, resulting in a maximum delay range $\Delta\tau_{\text{spherical-wave}}$. Following a similar extreme-value analysis, when the target distance is short or the array aperture is large, $\Delta\tau_{\text{spherical-wave}}-\Delta\tau_{\text{plane-wave}}=\frac{1}{\xi}(\gamma-\xi N)^2>0$, where $\gamma=-\frac{\delta\theta^{\prime}}{2c}>0$ and $\xi=\frac{\delta^2\alpha^{\prime}}{c}>0$~\cite{rainbow}. Therefore, compared with conventional wideband near-field beam training based on polar-domain codebooks, the proposed wideband near-field beam training scheme based on wavenumber-domain codebooks requires a smaller maximum TTD delay range.

\section{Simulation Results}
The simulation results presented in this section are intended to validate the superior performance of the proposed pattern zooming based wideband beam training method. We consider a wideband XL-MIMO system. Unless stated otherwise, the default system parameters are configured as follows. The Tx is equipped with $N=511$ antennas with an antenna spacing of $\delta=\lambda_\text{c} / 2$. The system operates at a central carrier frequency of $f_\text{c} = 30$ GHz and features a wide bandwidth of $B = 3$ GHz with $M=255$ subcarriers. The signal-to-noise ratio (SNR) is set to $20$ dB. The potential spatial angle range of the Tx is set to $\sin\theta \in [\sin(-\pi/3),\sin(\pi/3)]$ and the Rx distance from the Tx is randomly distributed within $[10\, \text{m}, 120\, \text{m}]$. The proposed scheme adopts a relative thresholding strategy with a threshold coefficient of $\tau=0.4$, which is determined through offline Monte Carlo calibration~\cite{dft_bt_2}. All reported numerical outcomes represent the statistical results obtained from over 1000 independent runs of Monte Carlo simulations.

\begin{figure*}[htbp]
	\centering
	\begin{minipage}[t]{0.32\linewidth}
		\centering
		\includegraphics[width=\linewidth]{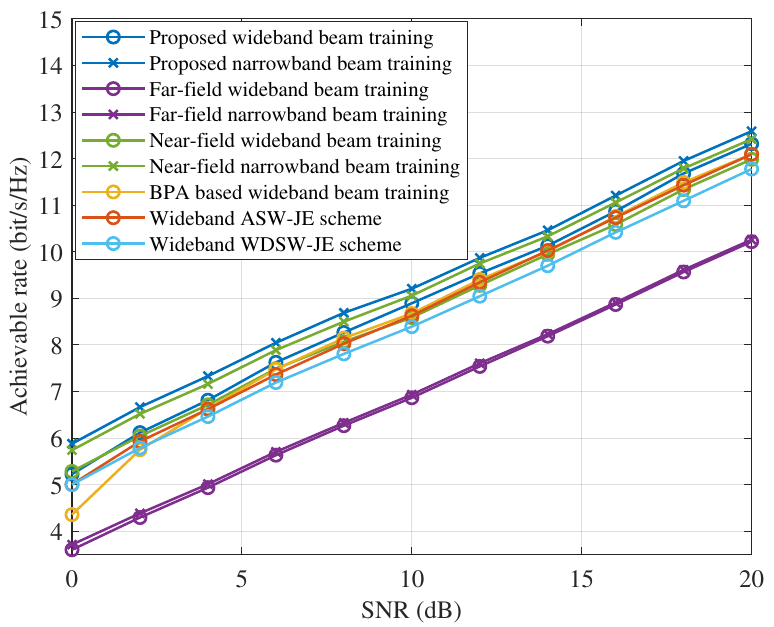}
		\caption{Achievable rate performance comparison with respect to the SNR.}
		\label{capacity_SNR}
	\end{minipage}\hfill
	\begin{minipage}[t]{0.32\linewidth}
		\centering
		\includegraphics[width=\linewidth]{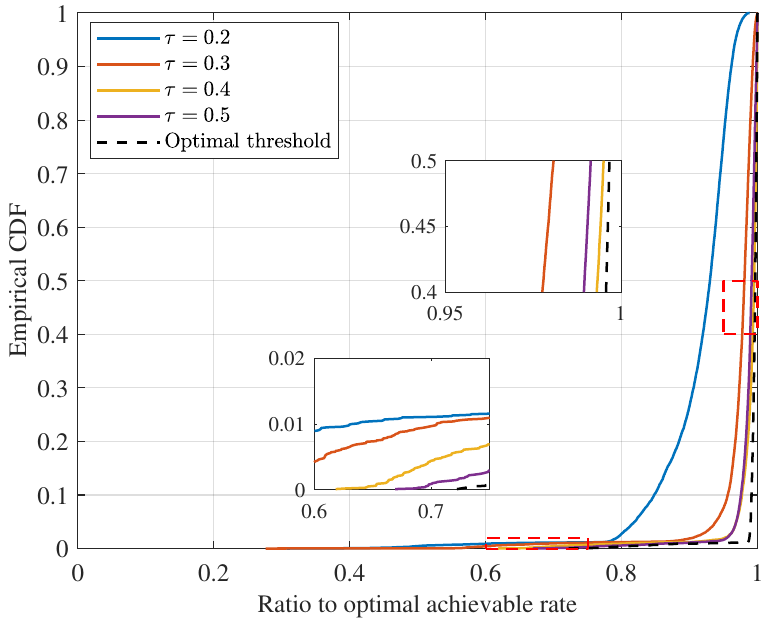}
		\caption{Empirical CDF of the ratio to optimal achievable rate.}
		\label{CDF}
	\end{minipage}\hfill
	\begin{minipage}[t]{0.32\textwidth}
		\centering
		\includegraphics[width=\linewidth]{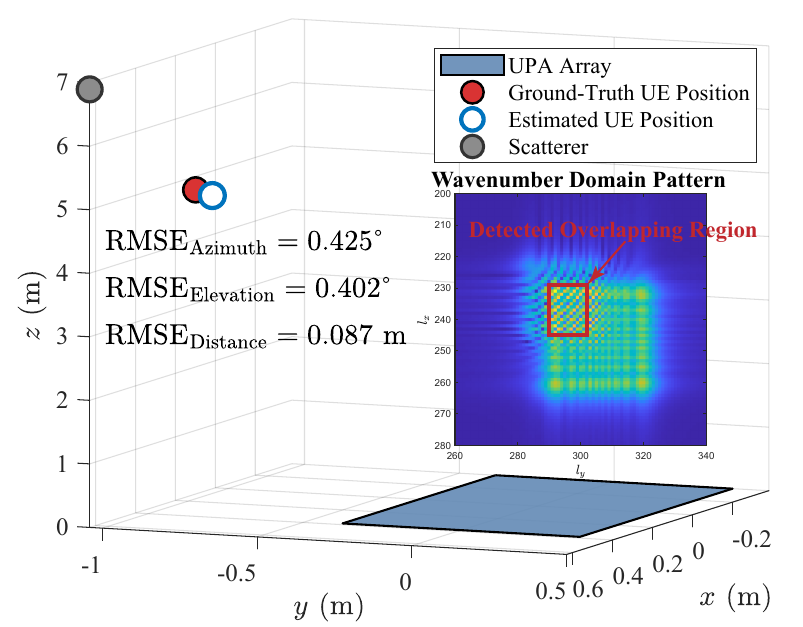}
		\caption{Position estimation performance under the UPA configuration with an NLoS scatterer.}
		\label{UPA_multipath}
	\end{minipage}\hfill
\end{figure*}

Regarding the antenna configuration, although the theoretical framework is developed for a general UPA, the numerical evaluations are conducted using a ULA setup. This simplification is motivated by the need for a direct and fair comparison with existing near-field beam training benchmarks\cite{dft_bt_1, dft_bt_2}, which are predominantly designed for ULA architectures. The ULA scenario is implemented simply by setting $l_y$ to zero in step 2 of the Algorithm \ref{alg1}. Simulation results under the general UPA configuration are presented in Fig.~\ref{UPA_multipath}, confirming the effectiveness of the proposed framework in 2D array scenarios.

For the purpose of comparative analysis, we consider the following schemes.

1) \textbf{Proposed wideband beam training}: This method utilizes controllable pattern zooming to efficiently acquire the wavenumber-domain pattern, enabling accurate estimation of the receiver's angle and distance.

2) \textbf{Proposed narrowband beam training}: This method estimates the receiver's angle and distance by acquiring the wavenumber-domain pattern through codebook-based sweeping performed exclusively on a single subcarrier.

3) \textbf{Far-field wideband beam training}: This method represents the classical wideband beam training for far-field communications\cite{wide_far}.

4) \textbf{Far-field narrowband beam training}: This method is the classical narrowband beam training designed for far-field scenarios\cite{narrow_far}.

5) \textbf{Near-field wideband beam training}: This method represents the classical near-field wideband beam training scheme\cite{rainbow}. The number of distance samples is set to $S=10$.

6) \textbf{Near-field narrowband beam training}: This method is the classical narrowband beam training designed for near-field scenarios with $S=10$ \cite{narrow_near}.

7) \textbf{Beam pattern analysis (BPA) based wideband beam training}: This method estimates the receiver location by fitting the beam pattern width and center to approximated closed-form expressions~\cite{dft_bt_2}.

8) \textbf{Wideband angle support width based joint angle and range estimation (ASW-JE) scheme}: This method estimates the user angle and range under DFT codebooks by determining the median of an angular support region and performing a 1D exhaustive search over a surrogate support width function~\cite{dft_bt_1}.

9) \textbf{Wideband wavenumber-domain support width based joint angle and range estimation (WDSW-JE) scheme}: This method inspects the near-field channel energy spectrum in the wavenumber domain, locating the user angle via spatial symmetry and resolving the range via substitution into a principle of stationary phase derived geometric width equation~\cite{dft_bt_5}.

To comprehensively evaluate the effectiveness of the proposed scheme, we employ three key performance metrics: the achievable rate, the root mean square error (RMSE) for angle and range estimation. Specifically, the achievable rate is defined as $R=\frac{1}{M} \sum_{m=1}^{M} \log_2 \left( 1 + \frac{P_t}{\sigma^2} \Vert \mathbf{h}_m \mathbf{v}_m \Vert ^2\right)$, where $\mathbf{v}_m$ is the beamforming vector for data transmission determined by different schemes and $P_t / \sigma^2$ denotes the SNR. Then, the RMSE for angle estimation is defined as $\text{RMSE}_{\text{angle}}= \sqrt{\mathbb{E}[(\theta_{\text{u}} - \hat{\theta}_{\text{u}})^2 ]}$, where $\theta_\text{u}$ and $\hat{\theta}_{\text{u}}$ denote the true and estimated receiver angle. Similarly, the RMSE for distance estimation is defined as $\text{RMSE}_{\text{distance}}= \sqrt{\mathbb{E}[(r_{\text{u}} - \hat{r}_{\text{u}})^2 ]}$, where $r_\text{u}$ and $\hat{r}_{\text{u}}$ denote the true and estimated receiver distance.

\begin{figure*}[htbp]
    \centering
    \begin{minipage}[t]{0.32\linewidth}
        \centering
        \includegraphics[width=\linewidth]{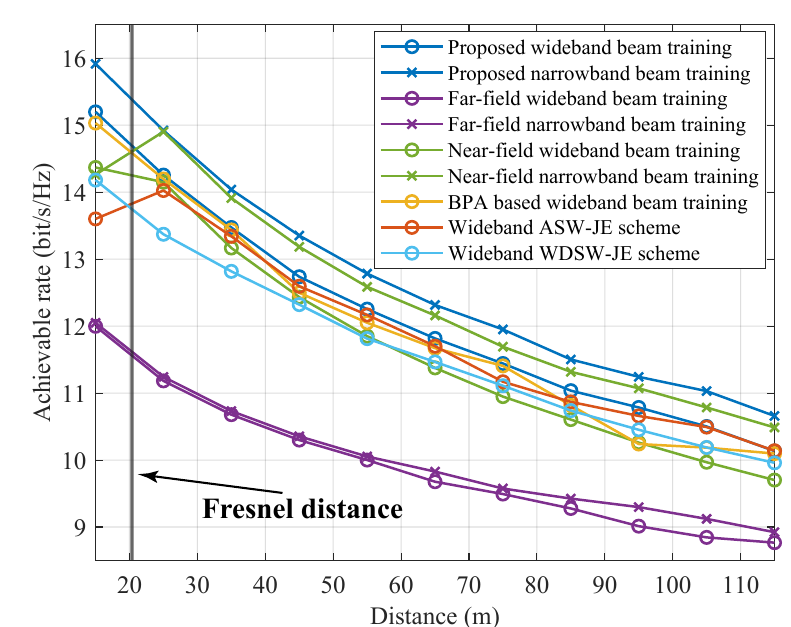}
        \caption{Achievable rate performance comparison with respect to the distance.}
        \label{capacity_distance}
    \end{minipage}\hfill
    \begin{minipage}[t]{0.32\linewidth}
        \centering
        \includegraphics[width=\linewidth]{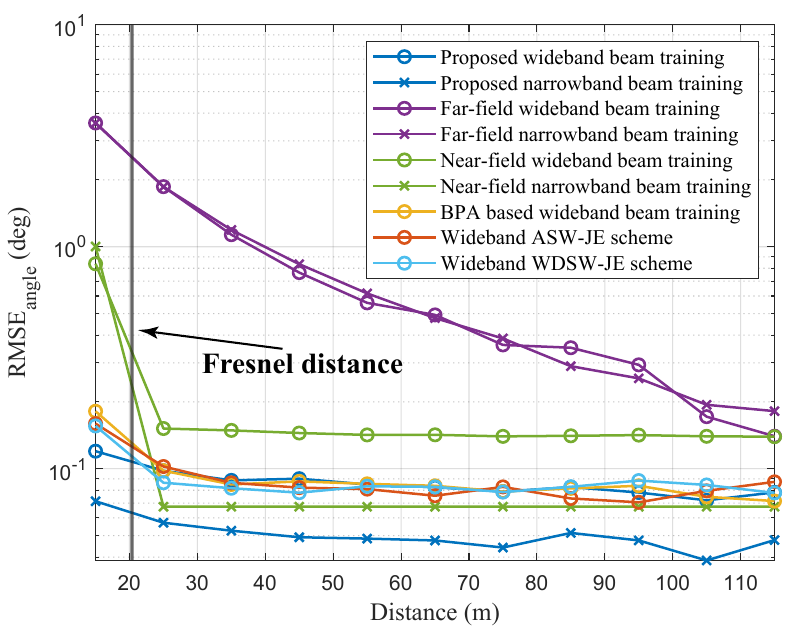}
        \caption{Angle estimation RMSE performance comparison with respect to the distance.}
        \label{RMSE_angle_distance}
    \end{minipage}\hfill
    \begin{minipage}[t]{0.32\textwidth}
        \centering
        \includegraphics[width=\linewidth]{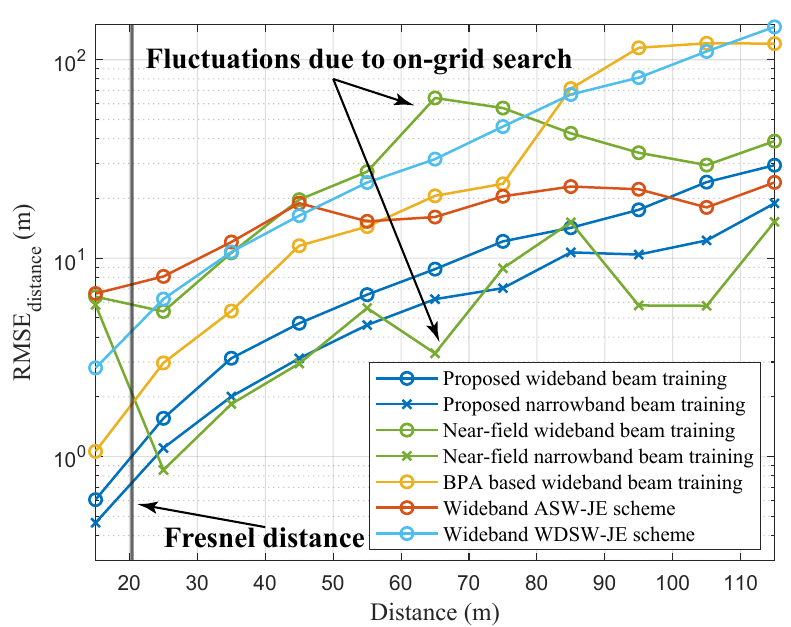}
        \caption{Distance estimation RMSE performance comparison with respect to the distance.}
        \label{RMSE_distance_distance}
    \end{minipage}\hfill
\end{figure*}

% \begin{figure}[htbp]
%     \centering
%     \includegraphics[width=8.5cm]{Figs/capacity_distance.pdf}
%     \caption{Achievable rate performance comparison with respect to the distance.} 
%     \label{capacity_distance}
% \end{figure}

\begin{figure*}[htbp]
	\centering
	\begin{minipage}[t]{0.32\linewidth}
		\centering
		\includegraphics[width=\linewidth]{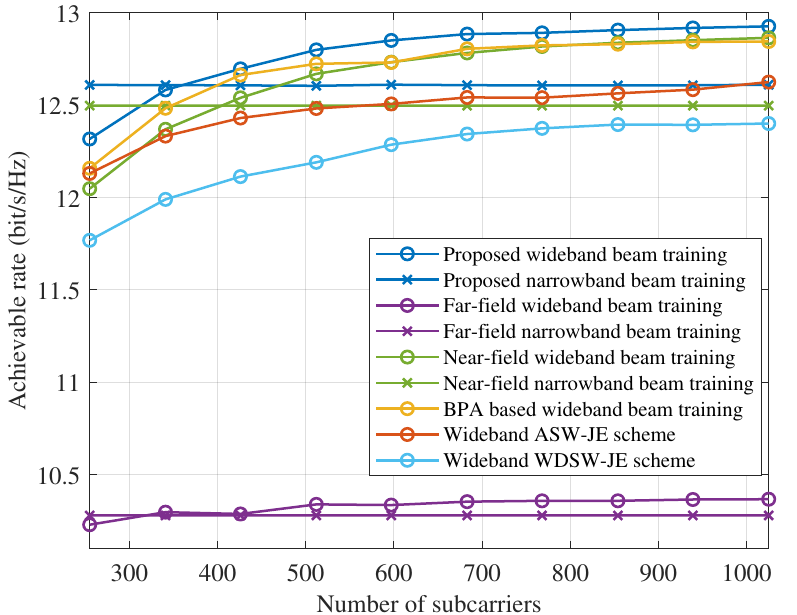}
		\caption{Achievable rate performance comparison with respect to the number of subcarriers.}
		\label{capacity_subcarrier}
	\end{minipage}\hfill
	\begin{minipage}[t]{0.32\linewidth}
		\centering
		\includegraphics[width=\linewidth]{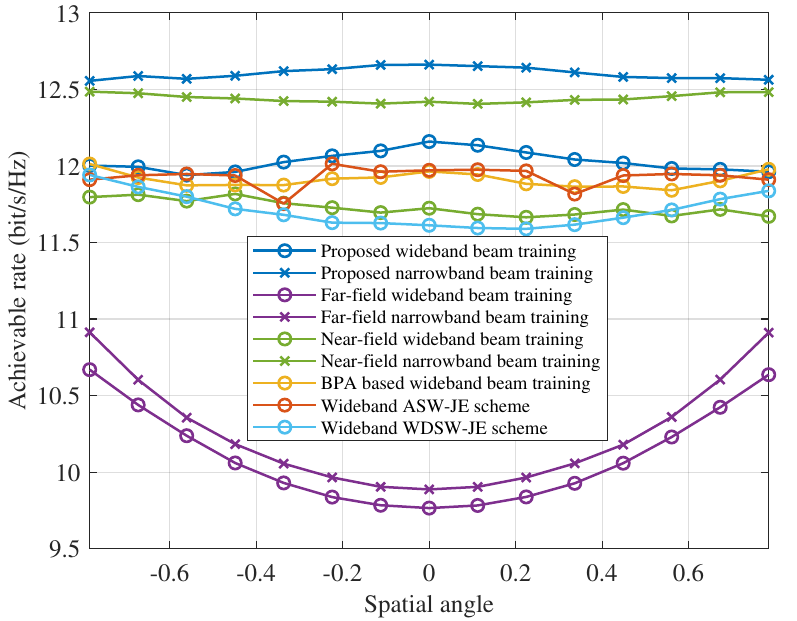}
		\caption{Achievable rate performance comparison with respect to the spatial angle.}
		\label{capacity_theta}
	\end{minipage}\hfill
	\begin{minipage}[t]{0.32\textwidth}
		\centering
		\includegraphics[width=\linewidth]{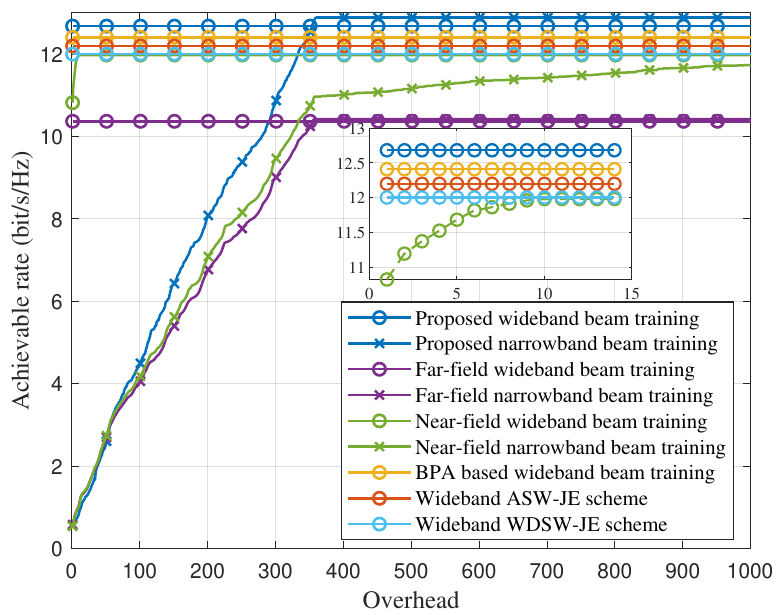}
		\caption{Achievable rate performance comparison with respect to the overhead.}
		\label{capacity_overhead}
	\end{minipage}\hfill
\end{figure*}

First of all, Fig. \ref{capacity_distance} compares the achievable rates of different methods against distances. Evidently, the proposed beam training method maintains superior performance across the examined distance range. Specifically, the proposed schemes and other alternative DFT codebook-based near-field beam training benchmarks, such as the BPA, ASW-JE, and WDSW-JE schemes, significantly outperform the traditional far-field and near-field benchmarks. Furthermore, a performance gain is observed for our proposed scheme over these DFT codebook-based near-field beam training benchmarks. This superiority stems from the approximation-free, precise analytical expressions for normalized wavenumber width and center derived in Section \uppercase\expandafter{\romannumeral4}-B. In contrast, the BPA, ASW-JE, and WDSW-JE schemes inevitably rely on approximated position-dependent functions. Moreover, for the traditional beam training methods, the far-field methods yield substantially lower rates than all other schemes, while the near-field methods experience a sharp performance drop as the distance is smaller than the Fresnel distance. These degradations are attributed to the reduced accuracy of the Taylor series expansion upon which these traditional approaches are based. The reason for the performance discrepancy between the same method in wideband and narrowband systems will be further explored in Fig. \ref{capacity_subcarrier}.

Fig. \ref{RMSE_angle_distance} compares the RMSE for angle estimation of different methods against distances. Examining the traditional far-field methods, it is evident that their inferior achievable rate performance in Fig. \ref{capacity_distance} stems not only from the absence of range information but also from their deteriorating angular estimation accuracy in the radiative near-field region. Similarly, the near-field benchmarks encounter a significant degradation in angular precision when the receiver is located within the reactive near-field. Regarding the proposed method and the alternative DFT codebook-based near-field beam training benchmarks, both exhibit comparable and superior angular estimation accuracy across most distance ranges. Although the proposed method employs exact expressions for angle estimation, it may exhibit slightly lower accuracy than certain benchmarks in specific scenarios. This is caused by practical discretization effects, where the fixed threshold coefficient and discretized pattern features introduce minor extraction errors.

Fig. \ref{RMSE_distance_distance} compares the RMSE for distance estimation of different methods against distances. As illustrated, our proposed methods achieve a significant leap in range estimation precision compared to these DFT codebook-based near-field beam training benchmarks, which is the primary source of the achievable rate gains observed in Fig. \ref{capacity_distance}. Regarding the performance trends, the estimation accuracy of both the proposed and DFT-based near-field benchmarks exhibits a gradual degradation as the communication distance increases. This is consistent with the fact that the normalized wavenumber width narrows at greater distances. Consequently, the finite resolution of the wavenumber-domain pattern leads to reduced estimation precision. However, it is worth noting that this upward RMSE trend at larger distances does not severely impact the achievable rate, as the system's sensitivity to distance errors diminishes as the receiver moves further into the far-field region. Most importantly, Fig. \ref{RMSE_distance_distance} highlights a distinct advantage of our off-grid estimation strategy. While the proposed and DFT-based near-field schemes maintain relatively smooth performance curves, the traditional on-grid range estimation methods exhibit severe fluctuations. These instabilities are inherent to on-grid searching, where performance is highly sensitive to whether the true receiver location aligns with the predefined grid points, whereas our off-grid approach ensures robust and stable localization.

In Fig. \ref{capacity_subcarrier}, we illustrate the achievable rate performance against the number of subcarriers. It is observed that the performance of all evaluated wideband beam training methods improves as the number of subcarriers $M$ increases, whereas the performance of their narrowband counterparts remains constant. This phenomenon is fundamentally dictated by the measurement precision of the spatial features, specifically the pattern width and center, which directly govern the accuracy of the estimated receiver coordinates. For narrowband schemes, the measurement precision relies on the normalized wavenumber resolution $2/N$. This resolution is constrained by the fixed number of antennas $N$, which explains why their performance curves remain flat regardless of the subcarrier quantity. Conversely, the wideband schemes probe a fixed wavenumber range across multiple subcarriers through pattern zooming. Therefore, deploying a larger number of subcarriers reduces the normalized wavenumber spacing between adjacent subcarriers. This densification enhances the measurement precision of the pattern width and center, leading to the growth in the achievable rate. Notably, when the number of subcarriers is relatively small, the narrowband methods exhibit superior performance due to the fine resolution provided by the massive antenna array. However, as $M$ continues to increase, the abundant subcarriers elevate the effective pattern resolution of the wideband system. This advantage enables the performance of the proposed wideband method to eventually cross over and surpass its narrowband counterpart.

Fig. \ref{capacity_theta} shows the achievable rate performance against the spatial angle $\theta_\text{u}$. As established in \cite{wideband_beamforming}, near-field characteristics manifest more prominently when the spatial angle is around zero. Consequently, the performance gain of our proposed wideband scheme over these benchmarks is most pronounced in this vicinity. This superiority stems from our method's enhanced range estimation accuracy, which is introduced in Fig. \ref{RMSE_distance_distance}. Similarly, far-field schemes, constrained by the absence of range information, suffer severe performance degradation at these angles. Moreover, an intriguing observation is the asymmetry in wideband performance, contrasting with the symmetric profiles of narrowband methods. This stems from their sweeping mechanisms: narrowband systems use a symmetric angular grid, while the spatial angles sampled across subcarriers in \eqref{Eq3_4} lack symmetry around the central angle, resulting in the observed asymmetric rate.

Fig. \ref{capacity_overhead} presents the achievable rate performance as a function of the training overhead. Here, an overhead of $n$ implies that the transmission beamforming vector is determined based on the feedback from $n$ beam measurements\cite{wch-twc}. It is evident that all wideband beam training methods require significantly lower overhead than their narrowband counterparts to achieve peak performance. This efficiency is attributed to the pattern zooming effect in wideband systems, where a single beam can probe multiple directions across different subcarriers. Furthermore, since we do not impose a constraint on the maximum delay range of the TTDs, the target angular range can be covered within a single measurement by selecting large periodic indices, as in \eqref{Eq3_4}. Even if practical constraints on the TTD delay range were introduced, the full angular space could still be explored through a limited number of additional tests. Thus, it should be emphasized that the training overhead required by existing near-field wideband beam training is $S-1$ times that of our proposed wideband beam training, rather than merely requiring $S-1$ additional beam measurements.

Fig. \ref{capacity_SNR} illustrates the achievable rate performance of various beam training schemes across a range of SNR, specifically from 6 dB to 20 dB. As expected, the achievable rate of all considered methods scales monotonically with the increase in SNR. Furthermore, the proposed wideband beam training scheme maintains a performance advantage over all benchmarks throughout the entire SNR regime. This performance gain remains robust even in low-SNR conditions, underscoring the effectiveness and reliability of the proposed methods.

Fig. \ref{CDF} illustrates the empirical cumulative distribution function (CDF) of the ratio between the achievable rate of the proposed scheme and the optimal performance upper bound, obtained under the assumption of perfect channel state information. Numerical results show that the threshold coefficient $\tau=0.4$ provides the best average performance among the evaluated fixed threshold coefficients. To further investigate the impact of threshold selection, an optimal threshold benchmark is included, where the threshold coefficient is optimally selected for each channel realization. Although the optimal threshold benchmark achieves the highest performance, its improvement over the fixed threshold coefficient is marginal. Moreover, the CDF curves for $\tau=0.3$, $\tau=0.4$, and $\tau=0.5$ almost overlap, indicating that the proposed algorithm is robust to moderate variations of the threshold coefficient.

To verify the derivations and evaluate the robustness of the proposed framework under planar architectures, we further conduct evaluations using a transmitter equipped with a UPA. To create a more challenging environment, an NLoS scattering path is introduced such that its wavenumber-domain pattern partially overlaps with the LoS pattern. As illustrated in Fig. ~\ref{UPA_multipath}, by executing the anomaly width detection mechanism via the multi-Otsu algorithm detailed in Section IV-D, the overlapping regions are accurately detected. Consequently, the proposed method successfully extracts the pattern width and center features from the non-overlapping regions. The results demonstrate that the proposed framework maintains high estimation accuracy for the azimuth angle, elevation angle, and communication distance even under multipath overlapping conditions with a transmitter UPA.

\section{Conclusion}
In this paper, we studied efficient near-field wideband beam training for XL-MIMO systems. To this end, we first developed a refined FPW representation to accurately characterizes the wideband channel across both far-field and near-field regions. Building on this model, we unveiled a novel pattern zooming effect in wideband system, which emerges when plane-wave transmission is employed across all field regions in wideband systems. Specifically, we revealed that wavenumber-domain patterns across different subcarriers undergo a frequency-dependent scaling relative to the center frequency, analogous to being viewed through a magnifying or reducing lens. Leveraging this effect, we propose an efficient near-field beam training method that configures TD units to simultaneously probe distinct wavenumber directions, enabling the rapid acquisition of the complete wavenumber-domain pattern. Furthermore, we derived exact, approximation-free closed-form expressions that establish a rigorous mapping between the receiver's spatial coordinates and the extracted scalar features, namely the pattern width and center. This robust feature extraction strategy enables the direct and precise estimation of the receiver's angle and distance. Finally, numerical results showed that the proposed method achieved higher achievable rate and lower training overhead compared to the benchmarks.

%\begin{appendices}
%\section{Derivation of $x_r$ \label{app_a}}
%From \eqref{Eq3_11}, we have
%\begin{equation}
%\label{Proof1_1}
%\begin{aligned}
%    &x_r=\frac{L_x}{2} \frac{(C_1+C_3-2)\pm 2\sqrt{(C_1-1)(C_3-1)}}{C_1-C_3}, \\
%    & = \frac{L_x}{2} \frac{\left(\sqrt{C_1-1}\pm \sqrt{C_3-1}\right)^2}{\left(\sqrt{C_1-1}+\sqrt{C_3-1}\right)\left(\sqrt{C_1-1}-\sqrt{C_3-1}\right)},
%\end{aligned}
%\end{equation}
%which simplifies to two distinct roots. Let $\mathcal{X} = \frac{\sqrt{C_1-1} -\sqrt{C_3-1}}{\sqrt{C_1-1} +\sqrt{C_3-1}}$. The solution corresponding to the minus sign in the numerator (i.e., $x_r = \frac{L_x}{2}\mathcal{X}$) yields values within the array aperture, satisfying $|x_r| \le L_x/2$. Conversely, the solution corresponding to the plus sign (i.e., $x_r = \frac{L_x}{2}\mathcal{X}^{-1}$) yields values outside the aperture. Additionally, for the boundary cases where $\vert x_r\vert = L_x/2$, either $C_1$ or $C_3$ approaches infinity, causing $\mathcal{X}$ to converge to $\pm 1$, which correctly yields $x_r = \pm L_x/2$. The introduction of the indicator function allows \eqref{Eq3_12} to compactly address all the various cases analyzed above.
%\end{appendices}

\bibliographystyle{IEEEtran}

\bibliography{ref}

\begin{thebibliography}{10}
\providecommand{\url}[1]{#1}
\csname url@samestyle\endcsname
\providecommand{\newblock}{\relax}
\providecommand{\bibinfo}[2]{#2}
\providecommand{\BIBentrySTDinterwordspacing}{\spaceskip=0pt\relax}
\providecommand{\BIBentryALTinterwordstretchfactor}{4}
\providecommand{\BIBentryALTinterwordspacing}{\spaceskip=\fontdimen2\font plus
\BIBentryALTinterwordstretchfactor\fontdimen3\font minus
  \fontdimen4\font\relax}
\providecommand{\BIBforeignlanguage}[2]{{%
\expandafter\ifx\csname l@#1\endcsname\relax
\typeout{** WARNING: IEEEtran.bst: No hyphenation pattern has been}%
\typeout{** loaded for the language `#1'. Using the pattern for}%
\typeout{** the default language instead.}%
\else
\language=\csname l@#1\endcsname
\fi
#2}}
\providecommand{\BIBdecl}{\relax}
\BIBdecl

\bibitem{6G-1}
Z.~Zhang, Y.~Xiao, Z.~Ma, M.~Xiao, Z.~Ding, X.~Lei, G.~K. Karagiannidis, and
  P.~Fan, ``{6G} wireless networks: Vision, requirements, architecture, and key
  technologies,'' \emph{IEEE Veh. Technol. Mag.}, vol.~14, no.~3, pp. 28--41,
  Sep. 2019.

\bibitem{NF-tutorial-1}
Y.~Liu, Z.~Wang, J.~Xu, C.~Ouyang, X.~Mu, and R.~Schober, ``Near-field
  communications: A tutorial review,'' \emph{IEEE Open J. Commun. Soc.},
  vol.~4, pp. 1999--2049, Aug. 2023.

\bibitem{mmWave-1}
Q.~Xue, C.~Ji, S.~Ma, J.~Guo, Y.~Xu, Q.~Chen, and W.~Zhang, ``A survey of beam
  management for mmwave and {THz} communications towards 6g,'' \emph{IEEE
  Commun. Surv. Tuts.}, vol.~26, no.~3, pp. 1520--1559, 3rd Quart. 2024.

\bibitem{NF-tutorial-2}
H.~Lu, Y.~Zeng, C.~You, Y.~Han, J.~Zhang, Z.~Wang, Z.~Dong, S.~Jin, C.-X. Wang,
  T.~Jiang \emph{et~al.}, ``A tutorial on near-field {XL-MIMO} communications
  toward {6G},'' \emph{IEEE Commun. Surv. Tuts.}, vol.~26, no.~4, pp.
  2213--2257, 4th Quart. 2024.

\bibitem{NF-tutorial-3}
Z.~Wang, J.~Zhang, H.~Du, D.~Niyato, S.~Cui, B.~Ai, M.~Debbah, K.~B. Letaief,
  and H.~V. Poor, ``A tutorial on extremely large-scale {MIMO} for {6G}:
  Fundamentals, signal processing, and applications,'' \emph{IEEE Commun. Surv.
  Tuts.}, vol.~26, no.~3, pp. 1560--1605, 3rd Quart. 2024.

\bibitem{NF-tutorial-4}
M.~Cui, Z.~Wu, Y.~Lu, X.~Wei, and L.~Dai, ``Near-field {MIMO} communications
  for {6G}: Fundamentals, challenges, potentials, and future directions,''
  \emph{IEEE Commun. Mag.}, vol.~61, no.~1, pp. 40--46, Jan. 2022.

\bibitem{Tcom-DLL}
M.~Cui and L.~Dai, ``Channel estimation for extremely large-scale {MIMO}:
  Far-field or near-field?'' \emph{IEEE Trans. Commun.}, vol.~70, no.~4, pp.
  2663--2677, Apr. 2022.

\bibitem{rainbow}
M.~Cui, L.~Dai, Z.~Wang, S.~Zhou, and N.~Ge, ``Near-field rainbow: Wideband
  beam training for {XL-MIMO},'' \emph{IEEE Trans. Wireless Commun.}, vol.~22,
  no.~6, pp. 3899--3912, Jun. 2022.

\bibitem{wideband_beamforming}
M.~Cui and L.~Dai, ``Near-field wideband beamforming for extremely large
  antenna arrays,'' \emph{IEEE Trans. Wireless Commun.}, vol.~23, no.~10, pp.
  13\,110--13\,124, Oct. 2024.

\bibitem{wide_beamforming_lyw}
Z.~Wang, X.~Mu, and Y.~Liu, ``Beamfocusing optimization for near-field wideband
  multi-user communications,'' \emph{IEEE Trans. Commun.}, vol.~73, no.~1, pp.
  555--572, Jan. 2024.

\bibitem{Rayleigh}
K.~T. Selvan and R.~Janaswamy, ``Fraunhofer and {Fresnel} distances: Unified
  derivation for aperture antennas,'' \emph{IEEE Antennas Propag. Mag.},
  vol.~59, no.~4, pp. 12--15, Aug. 2017.

\bibitem{boundary_3}
S.~Sun, R.~Li, C.~Han, X.~Liu, L.~Xue, and M.~Tao, ``How to differentiate
  between near field and far field: Revisiting the {Rayleigh} distance,''
  \emph{IEEE Commun. Mag.}, vol.~63, no.~1, pp. 22--28, Jan. 2025.

\bibitem{wch-twc}
C.~Weng, X.~Guo, Y.~Guo, and Y.~Wang, ``Wavenumber domain beam training in
  {XL-MIMO} systems: Unifying far-field and near-field,'' \emph{IEEE Trans.
  Wireless Commun.}, vol.~25, pp. 2183--2196, Aug. 2025.

\bibitem{wideband_bt}
T.~Zheng, M.~Cui, Z.~Wu, and L.~Dai, ``Near-field wideband beam training based
  on distance-dependent beam split,'' \emph{IEEE Trans. Wireless Commun.},
  vol.~24, no.~2, pp. 1278--1292, Feb. 2024.

\bibitem{far-hierarchical}
C.~Qi, K.~Chen, O.~A. Dobre, and G.~Y. Li, ``Hierarchical codebook-based
  multiuser beam training for millimeter wave massive {MIMO},'' \emph{IEEE
  Trans Wireless Commun.}, vol.~19, no.~12, pp. 8142--8152, Dec. 2020.

\bibitem{narrow_near}
Z.~Wu and L.~Dai, ``Multiple access for near-field communications: {SDMA} or
  {LDMA}?'' \emph{IEEE J. Sel. Areas Commun.}, vol.~41, no.~6, pp. 1918--1935,
  Jun. 2023.

\bibitem{near-hierarchical-1}
C.~Wu, C.~You, Y.~Liu, L.~Chen, and S.~Shi, ``Two-stage hierarchical beam
  training for near-field communications,'' \emph{IEEE Trans. Veh. Technol.},
  vol.~73, no.~2, pp. 2032--2044, Feb. 2023.

\bibitem{wch-TVT}
C.~Weng, X.~Guo, and Y.~Wang, ``Near-field beam training with hierarchical
  codebook: Two-stage learning-based approach,'' \emph{IEEE Trans. Veh.
  Technol.}, vol.~73, no.~9, pp. 14\,003--14\,008, Sep. 2024.

\bibitem{near-hierarchical-2}
Y.~Lu, Z.~Zhang, and L.~Dai, ``Hierarchical beam training for extremely
  large-scale {MIMO}: From far-field to near-field,'' \emph{IEEE Trans.
  Commun.}, vol.~72, no.~4, pp. 2247--2259, Apr. 2023.

\bibitem{learning-1}
W.~Liu, H.~Ren, C.~Pan, and J.~Wang, ``Deep learning based beam training for
  extremely large-scale massive {MIMO} in near-field domain,'' \emph{IEEE
  Commun. Lett.}, vol.~27, no.~1, pp. 170--174, Jan. 2022.

\bibitem{dft_bt_4}
H.~Heo and W.~Choi, ``{DFT}-based near-field beam alignment: Model-based and
  data-driven hybrid approach,'' \emph{IEEE Trans. Wireless Commun.}, vol.~25,
  pp. 11\,850--11\,865, Feb. 2026.

\bibitem{dft_bt_2}
Z.~Wang, R.~Kiran, S.~Tsai, and R.~Zhang, ``Low-complexity near-field beam
  training with {DFT} codebook based on beam pattern analysis,'' \emph{arXiv
  preprint arXiv:2503.21954}, 2025.

\bibitem{dft_bt_3}
Z.~Wang, S.~Tsai, R.~Kiran, and R.~Zhang, ``Precise near-field beam training
  with {DFT} codebook based on amplitude-only measurement,'' \emph{IEEE Trans.
  Wireless Commun.}, vol.~25, pp. 14\,503--14\,516, Apr. 2026.

\bibitem{dft_bt_1}
X.~Wu, C.~You, J.~Li, and Y.~Zhang, ``Near-field beam training: Joint angle and
  range estimation with {DFT} codebook,'' \emph{IEEE Trans. Wireless Commun.},
  vol.~23, no.~9, pp. 11\,890--11\,903, Sep. 2024.

\bibitem{wideband_PS}
F.~Sohrabi and W.~Yu, ``Hybrid analog and digital beamforming for {mmWave OFDM}
  large-scale antenna arrays,'' \emph{IEEE J. Sel. Areas Commun.}, vol.~35,
  no.~7, pp. 1432--1443, Jul. 2017.

\bibitem{split_1}
X.~Gao, L.~Dai, S.~Zhou, A.~M. Sayeed, and L.~Hanzo, ``Wideband beamspace
  channel estimation for millimeter-wave mimo systems relying on lens antenna
  arrays,'' \emph{IEEE Trans. Signal Process.}, vol.~67, no.~18, pp.
  4809--4824, Sep. 2019.

\bibitem{split_2}
J.~Tan and L.~Dai, ``Wideband beam tracking in {THz} massive mimo systems,''
  \emph{IEEE J. Sel. Areas Commun.}, vol.~39, no.~6, pp. 1693--1710, Jun. 2021.

\bibitem{wide_far}
V.~Boljanovic, H.~Yan, C.-C. Lin, S.~Mohapatra, D.~Heo, S.~Gupta, and
  D.~Cabric, ``Fast beam training with true-time-delay arrays in wideband
  millimeter-wave systems,'' \emph{IEEE Trans Circuits and Syst. I, Reg.
  Papers}, vol.~68, no.~4, pp. 1727--1739, Apr. 2021.

\bibitem{wd_codebook}
J.~Yang, Y.~Chen, Y.~Si, H.~Yu, Y.~Sun, S.~Zhang, and Z.~Lu, ``Revealing the
  evanescent components in kronecker product-based codebooks: Insights and
  applications,'' \emph{IEEE Trans. Wireless Commun.}, 2025, early access.

\bibitem{MIMO_DLL}
Y.~Lu and L.~Dai, ``Near-field channel estimation in mixed {LoS/NLoS}
  environments for extremely large-scale {MIMO} systems,'' \emph{IEEE Trans.
  Commun.}, vol.~71, no.~6, pp. 3694--3707, Jun. 2023.

\bibitem{antenna_deactivation}
Z.~Xiao, T.~He, P.~Xia, and X.-G. Xia, ``Hierarchical codebook design for
  beamforming training in millimeter-wave communication,'' \emph{IEEE Trans.
  Wireless Commun.}, vol.~15, no.~5, pp. 3380--3392, May 2016.

\bibitem{pizzo-1}
A.~Pizzo, L.~Sanguinetti, and T.~L. Marzetta, ``Fourier plane-wave series
  expansion for holographic {MIMO} communications,'' \emph{IEEE Trans. Wireless
  Commun.}, vol.~21, no.~9, pp. 6890--6905, Sep. 2022.

\bibitem{ellipse}
X.~Guo, Y.~Chen, Y.~Wang, Z.~Wang, and C.~Yuen, ``Wavenumber-domain near-field
  channel estimation: Beyond the fresnel bound,'' in \emph{GLOBECOM 2024-2024
  IEEE Global Communications Conference}.\hskip 1em plus 0.5em minus
  0.4em\relax IEEE, 2024, pp. 4660--4665.

\bibitem{Nyquist_pizzo}
A.~Pizzo, A.~de~Jesus~Torres, L.~Sanguinetti, and T.~L. Marzetta, ``Nyquist
  sampling and degrees of freedom of electromagnetic fields,'' \emph{IEEE
  Trans. Signal Process.}, vol.~70, pp. 3935--3947, Jun. 2022.

\bibitem{Full-TTD}
M.~Monemi, M.~A. Fallah, M.~Rasti, O.~Yazdani, O.~L. Lopez, and M.~Latva-aho,
  ``Beam squint mitigation in wideband hybrid beamformers: {Full-TTD},
  {sparse-TTD}, or {non-TTD}?'' \emph{IEEE Wireless Commun.}, 2026, early
  access.

\bibitem{dft_bt_5}
H.~Xing, Y.~Zhang, J.~Zhang, H.~Xu, G.~Liu, and Q.~Wang, ``An approximate
  wave-number domain expression for near-field {XL}-array channel,'' \emph{IEEE
  Trans. Veh. Technol.}, vol.~74, no.~5, pp. 8267--8272, May 2025.

\bibitem{multipath}
Z.~Wang, A.~M. Nivetha, Y.~Hu, and R.~Zhang, ``Compressive beam-pattern-aware
  near-field beam training via total variation denoising,'' \emph{IEEE Wireless
  Commun. Lett.}, vol.~15, pp. 2914--2918, May 2026.

\bibitem{wd_channel}
Y.~Guo, Y.~Chen, and Y.~Wang, ``Channel estimation for holographic {MIMO}:
  Wavenumber-domain sparsity-inspired approaches,'' \emph{IEEE Wireless Commun.
  Lett.}, vol.~13, no.~8, pp. 2305--2309, Aug. 2024.

\bibitem{multi-Otsu}
P.-S. Liao, T.-S. Chen, P.-C. Chung \emph{et~al.}, ``A fast algorithm for
  multilevel thresholding,'' \emph{J. Inf. Sci. Eng.}, vol.~17, no.~5, pp.
  713--727, 2001.

\bibitem{narrow_far}
S.~Noh, M.~D. Zoltowski, and D.~J. Love, ``Multi-resolution codebook and
  adaptive beamforming sequence design for millimeter wave beam alignment,''
  \emph{IEEE Trans. Wireless Commun.}, vol.~16, no.~9, pp. 5689--5701, Sep.
  2017.

\end{thebibliography}
\newpage

\vfill

\end{document}